\documentclass[11pt]{article}
\usepackage{amssymb,amscd,array}
\catcode `\@=11
\@addtoreset{equation}{subsection}

\def\harr#1#2{\smash{\mathop{\hbox to .5in{\rightarrowfill}}
\limits^{\scriptstyle#1}_{\scriptstyle#2}}}

\def\harrl#1#2{\smash{\mathop{\hbox to .5in{\leftarrowfill}}
\limits^{\scriptstyle#1}_{\scriptstyle#2}}}

\def\qed{\blacksquare}
\newcommand{\be}{\begin{equation}}
\newcommand{\ee}{\end{equation}}
\newcommand{\bea}{\begin{eqnarray}}
\newcommand{\eea}{\end{eqnarray}}
\newcommand{\R}{\mathbb{R}}

\newcommand{\C}{\mathbb{C}}
\newtheorem{thm}{Theorem}[section]
\newtheorem{rem}[thm]{Remark}

\newtheorem{cor}[thm]{Corollary}
\newtheorem{prop}[thm]{Proposition}
\begin{document}
\begin{titlepage}
%\thispagestyle{empty}
%\begin{flushright}
%IFA-FT-402-1994, November
%\end{flushright}
%\bigskip\bigskip
\begin{center}
{\bf \Large{The Standard Model and its Generalizations ~\\
in Epstein-Glaser Approach to Renormalization Theory II: the Fermion Sector and
the Axial Anomaly\\}}
\end{center}
\vskip 1.0truecm
\centerline{D. R. Grigore
\footnote{e-mail: grigore@theor1.theory.nipne.ro, grigore@theory.ifa.ro}}
\vskip5mm
\centerline{Dept. of Theor. Phys., Inst. Atomic Phys.}
\centerline{Bucharest-M\u agurele, P. O. Box MG 6, ROM\^ANIA}
\vskip 2cm
\bigskip \nopagebreak
\begin{abstract}
\noindent
We complete our study of non-Abelian gauge theories in the framework of
Epstein-Glaser approach to renormalization theory including in the model an
arbitrary number of Dirac Fermions. We consider the consistency of the model up
to the third order of the perturbation theory. In the second order we obtain
pure group theoretical relations expressing a representation property of the
numerical coefficients appearing in the left and right handed components of the
interaction Lagrangian. In the third order of the perturbation theory we obtain
the the condition of cancellation of the axial anomaly.
\end{abstract}
%\newpage\setcounter{page}1 
\end{titlepage}
                        
\section{Introduction}
                        
In some preceding papers \cite{Gr1}, \cite{Gr2} we have extended results of
Aste, D\"utsch and Scharf \cite{AS}, \cite{DS}, \cite{ASD3} concerning the
uniqueness of the non-Abelian gauge theory describing the consistent
interaction of Bosons of spin $1$. It appeared that the gauge invariance
principle is a natural consequence of the description of spin-one particles in
a factor Hilbert space: gauge invariance expresses the possibility of
factorising the $S$-matrix to the physical space, which is usually constructed
using the existence of a supercharge $Q$ according to the cohomological-type
formula:
$
{\cal H}_{phys} = Ker(Q)/Im(Q).  
$ 
The obstructions to such a factorization process are the well-known anomalies.
The case when the spin-one Bosons of non-null mass are admitted in the game was
studied in \cite{DS}, \cite{ASD3} for the concrete case of the electro-weak
interaction i.e. when the gauge group is exactly 
$SU(2) \times U(1)$.
                   
In \cite{Gr2} we have analysed the same problem considering that the spin-one
Bosons can have non-null masses and we did not impose any restriction on their
number and masses and we did not took into account the matter fields. Similar
results have been obtained in \cite{Sc2}. We have obtained only from the
condition of absence of the anomaly up to the second order the existence of a
Lie algebra $\mathfrak{g}$ and the existence of a representation of this Lie
algebra pertaining to the Higgs fields.
                        
In this paper, we consider the effect of including Dirac Fermions. In this way
we are able to investigate a truly realistic model of gauge interactions of
elementary particle and, in particular, to see what are the restrictions on
such a model determined by the cancellation of all anomalies. The main results
are the following ones.
                        
(A) The cancellation of the anomaly in the second order of the perturbation
theory brings new relations on the numerical coefficients of the left and right
handed components of the interaction Lagrangian. More precisely, new group
theoretical property appear:

(i) The coefficients of the vectorial and pseudo-vectorial couplings can be
organised as two representations of the gauge algebra:
$t^{+}_{a}$ 
and
$t^{-}_{a}$ 
with 
$a, b, \dots = 1,\dots,r$ 
group indices; the usual notations are 
$t^{R}_{a}$ 
and 
$t^{L}_{a}$.

(ii) The coefficients of the scalar and pseudo-scalar couplings can be
organised as some tensor operators. 

Some of these relations have been obtained from different considerations in
\cite{We1}, \cite{BBBC}.
                        
(B) The cancellation of the anomaly in the third order of the perturbation
theory gives, essentially, the usual condition of cancellation of the axial
anomaly:
\be
A_{abc} \equiv
Tr\left(t^{+}_{a} \{t^{+}_{b},t^{+}_{c}\}\right)
- Tr\left(t^{-}_{a} \{t^{-}_{b},t^{-}_{c}\}\right).
\label{ABBJ}
\ee

This is the expression of the Adler-Bardeen-Bell-Jackiw anomaly \cite{Ad},
\cite{BJ}, \cite{AB}, \cite{Ba}, \cite{Sch1}, \cite{LS1}, \cite{LS2},
\cite{We}.

Some conditions on the couplings of the Higgs fields appear if one imposes the
additional requirement that no finite renormalizations of degree greater than
$4$ are allowed. This conditions gives the usual expression for the Higgs
potential \cite{DS}, \cite{ASD3} for the case of the Standard Model (SM). In
\cite{Sc2} and \cite{DS1} the analysis of the pure Boson sector is performed
somewhat differently. We will comment on that in the next Section.
                        
The structure of the paper is the following one. In the next Section we
define the model and construct the interaction Lagrangian including Dirac
Fermions also. Then in Section \ref{perturbation} we outline the general
setting for the study of the renormalization theory, the general structure of
Ward identities and some facts about distribution splitting. In Section
\ref{second} we construct the $S$-matrix up to the second order of the
perturbation theory. For the case without matter fields we reobtain the
results of \cite{Gr2}. Then we consider the coupling of Yang-Mills fields with
Dirac Fermions and, as anticipated above, we obtain some interesting
group-theoretical relations; their complete analysis could fix very severely
the possible generalisations of the standard model. We also analyse the
conservation of the BRST current in the second order of the perturbation
theory. In Section \ref{third} we go to the third order of the perturbation
theory. We investigate the Dirac Fermionic sector and we get the new conditions
on the Fermionic representations (a) or (b) from above. In the end we
particularise the formalism for the case of the standard model with one
generation of Dirac particles. In the last Section we comment on the
possibility of removing the anomalies in all orders of perturbation theory.

For the sake of clarity of the rather long and intricate analysis we adopt the
mathematical {\it definition - theorem} style of presenting various assertions
and computations. 
\newpage

\section{General Description of the Vector Bosons\label{vectorbosons}}

\subsection{Massive Yang-Mills Fields\label{ym}}
         
In \cite{Gr1} and \cite{Gr2} we have started from the following two facts:

(1) a system of free zero-mass vector Bosons can be described in a Hilbert
space generated from the vacuum
$\Omega$
by applying the free fields
$
A_{\mu},~ u,~\tilde{u}
$
of zero mass and a factorization procedure induced by a supercharge operator;

(2) a system of free vector Bosons of mass
$m > 0$
can be described in a Hilbert space generated from the vacuum
$\Omega$
by applying the free fields
$
A_{\mu},~ u,~\tilde{u},~\Phi
$
of mass
$m$ and a factorization procedure induced by a supercharge operator.

Here
$A_{\mu}$
is a Boson vector field,
$u$
and
$\tilde{u}$
are scalar Fermi fields and
$\Phi$
is a scalar Boson fields; the fields
$u,~ \tilde{u}, \Phi$
are usually called ghost fields.

For the Yang-Mills model we combine somehow these two cases. We consider the 
auxiliary Hilbert space 
${\cal H}_{YM}^{gh,r}$
generated from the vacuum
$\Omega$
by applying the free fields
$
A_{a\mu},~ u_{a},~\tilde{u}_{a},~\Phi_{a} \quad a = 1,\dots,r
$
where the first one has vector transformation properties with respect to the
Poincar\'e group and the others are scalars. In other words, every vector field
has three scalar partners. Also
$
u_{a},~\tilde{u}_{a} \quad a = 1,\dots,r
$
are Fermion and
$
A_{\mu},~\Phi_{a} \quad a = 1,\dots,r
$
are Boson fields. 

We have two distinct possibilities for distinct indices $a$:

(I) Fields of type I correspond to an index $a$ such that the vector field
$A_{a}^{\mu}$ 
has non-zero mass
$m_{a}$.
In this case we suppose that all the other scalar partners fields
$
u_{a},~\tilde{u}_{a},~\Phi_{a}
$
have the same mass $m_{a}$.

(II) Fields of type II correspond to an index $a$ such that the vector field
$A_{a}^{\mu}$ 
has zero mass. In this case we suppose that the scalar partners fields
$
u_{a},~\tilde{u}_{a}
$
also have the zero mass but the scalar field 
$\Phi_{a}$
can have a non-zero mass: 
$m^{H}_{a} \geq 0$. 
It is convenient to use the compact notation 
\be
m^{*}_{a} \equiv \left\{\begin{array}{rcl} 
m_{a} & \mbox{for} & m_{a} \not= 0 \\
m^{H}_{a} & \mbox{for} & m_{a} = 0\end{array}\right. 
\ee

Then the following following equations of motion describe the preceding
construction: 
\be
(\square + m_{a}^{2}) u_{a}(x) = 0, \quad 
(\square + m_{a}^{2}) \tilde{u}_{a}(x) = 0, \quad
(\square + (m^{*}_{a})^{2}) \Phi_{a}(x) = 0, \quad a = 1,\dots,r.
\label{equ-r}
\ee

We also postulate the following canonical (anti)commutation relations:
\bea
\left[A_{a\mu}(x),A_{b\nu}(y)\right] = - 
\delta_{ab} g_{\mu\nu} D_{m_{a}}(x-y) \times {\bf 1},
\nonumber \\
\{u_{a}(x),\tilde{u}_{b}(y)\} = \delta_{ab} D_{m_{a}}(x-y) \times {\bf 1}, 
\quad
[ \Phi_{a}(x),\Phi_{b}(y) ] = \delta_{ab} D_{m^{*}_{a}}(x-y) \times {\bf 1};
\label{comm-r}
\eea
all other (anti)commutators are null.  
                        
In this Hilbert space we suppose given a sesquilinear form 
$<\cdot, \cdot>$
such that:
\be
A_{a\mu}(x)^{\dagger} = A_{a\mu}(x), \quad
u_{a}(x)^{\dagger} = u_{a}(x), \quad
\tilde{u}_{a}(x)^{\dagger} = - \tilde{u}_{a}(x), \quad
\Phi_{a}(x)^{\dagger} = \Phi_{a}(x).
\label{conjugate-YM}
\ee
                        
The ghost degree is $1$ for the fields
$
~u_{a},~\tilde{u}_{a}, \quad a = 1,\dots,r
$
and $0$ for the other fields.

One can define the BRST {\it supercharge} $Q$ by:
\bea
\{Q, u_{a} \}= 0 \quad 
\{Q, \tilde{u}_{a} \}= - i (\partial_{\mu} A_{a}^{\mu} + m_{a} \Phi_{a})
\nonumber \\
\{Q, A_{a}^{\mu} \} = i \partial^{\mu} u_{a} \quad
\{Q, \Phi_{a} \} = i m_{a} u_{a}, \quad \forall a = 1,\dots,r
\label{BRST-YM}
\eea
and 
\be
Q \Omega = 0.
\ee

Then one can justify that the {\bf physical} Hilbert space of the Yang-Mills
system is a factor space
\be
{\cal H}^{r}_{YM} \equiv {\cal H} \equiv Ker(Q)/Ran(Q).
\ee

The sesquilinear form
$<\cdot, \cdot>$
induces a {\it bona fide} scalar product on the Hilbert factor space.

The factorization process leads to the following {\bf physical} particle
content of this model:
\begin{itemize}
\item
For 
$m_{a} > 0$
the fields
$
A_{a}^{\mu},~u_{a},~\tilde{u}_{a},~\Phi_{a}
$
describe a particle of mass
$m_{a} > 0$
and spin $1$; this are the so-called {\it heavy Bosons} \cite{Gr2}.
\item
For 
$m_{a} = 0$
the fields
$
A_{a}^{\mu},~u_{a},~\tilde{u}_{a}
$
describe a particle of mass
$0$
and helicity $1$; the typical example is the {\it photon} \cite{Gr1}.
\item
For 
$m_{a} = 0$
the fields
$
\Phi_{a}
$
describe a scalar fields of mass
$m_{a}^{H}$;
this are the so-called {\it Higgs fields}.
\end{itemize}

This framework is sufficient for the study of the Standard Model (SM) of the
electro-weak interactions: indeed one takes
$r = 4$
and considers that there are three fields of type I and one field of type II.
The scalar field appearing in the last case can be considered as the Higgs
field.  

To include also quantum chromodynamics one must consider that there is a third
case:

(III) Fields of type III correspond to an index $a$ such that the vector field
$A_{a}^{\mu}$ 
has zero mass, the scalar partners
$
u_{a},~\tilde{u}_{a}
$
also have zero mass but the scalar field 
$\Phi_{a}$
is absent. 

In \cite{Sc2} and \cite{DS1} the model is constructed somewhat differently: one
eliminates the fields of type II and includes a number of supplementary scalar 
Bosonic fields
$\varphi_{i}$
of masses
$m_{i} \geq 0$.
In this framework one can consider for instance the very interesting
Higgs-Kibble model in which there are no zero-mass particle, so the adiabatic
limit probably exists.

One can preserve the general framework with only two types of indices if we
consider that in case II there are in fact three subcases (i.e three types of
indices $a$ for which 
$m_{a} = 0$):

(IIa) In this case
$A_{a\mu},~ u_{a},~ \tilde{u}_{a},~ \Phi_{a} \not\equiv 0$;

(IIb) In this case
$\Phi_{a} \equiv 0$;

(IIc) In this case
$A_{a\mu},~ u_{a},~ \tilde{u}_{a} \equiv 0$.

One must modify appropriately the canonical (anti)commutation  relations
(\ref{comm-r}) to avoid contradiction for some values of the indices. 
One has some freedom of notation: for instance, one can eliminate case (IIa)
if one includes the first three fields fields in case (IIb) and the last one in
case (IIc). The relations (\ref{BRST-YM}) are not affected in this way.

Let us consider the set of Wick monomials ${\cal W}$ constructed from the free
fields
$A_{a}^{\mu},~u_{a},~\tilde{u}_{a}$
and
$\Phi_{a}$
for all indices
$a = 1,\dots,r$;
we define the BRST operator
$d_{Q}: {\cal W} \rightarrow {\cal W}$
as the (graded) commutator with the supercharge operator $Q$. Then one can
prove easily that:
\be
d_{Q}^{2} = 0.
\label{coomology}
\ee

The class of observables on the factor space is defined as follows: an operator
$O: {\cal H}_{YM}^{gh,r} \rightarrow {\cal H}_{YM}^{gh,r}$
induces a well defined operator 
$[O]$
on the factor space
$\overline{Ker(Q)/Im(Q)} \simeq {\cal F}_{m}$
if and only if it verifies:
$
\left. d_{Q} O \right|_{Ker(Q)} = 0.
\label{dQ}
$
Because of the relation (\ref{coomology}) not all operators verifying the
condition (\ref{dQ}) are interesting. In fact, the operators of the type 
$d_{Q}O$ are inducing a null operator on the factor space; explicitly, we have:
\be
[d_{Q} O] = 0.
\ee

The canonical dimension
$\omega(W)$
of certain Wick monomial is defined according to the usual prescription.
By definition, a {\it Wick polynomial} is a sum of Wick monomials.

We will construct a perturbation theory {\it \'a la} Epstein-Glaser using this
set of free fields and imposing the usual axioms of causality, unitarity and
relativistic invariance on the chronological products
$T(x_{1},\dots,x_{n})$.
Moreover, we want that the result factorizes to the physical Hilbert space in
the adiabatic limit. This amounts to
\be
lim_{\epsilon \searrow 0} 
\left. d_{Q} \int_{(\R^{4})^{\times n}} dx_{1} \cdots dx_{n}
g_{\epsilon}(x_{1}) \cdots g_{\epsilon}(x_{n}) 
T(x_{1},\dots,x_{n}) \right|_{Ker(Q)} = 0, 
\quad \forall n \geq 1.
\label{gi-Tn}
\ee
                        
If this condition if fulfilled, then the chronological and the
antichronological products do factorise to the physical Hilbert space and they
give a perturbation theory verifying causality, unitarity and relativistic
invariance.

One may raise at this point the rather serious objection that the adiabatic
limit probably does not exists. One way to ``cure" this problem is to replace
the condition of factorisation (\ref{gi-Tn}) by the ``infinitesimal" version
postulated in \cite{AS} - \cite{DS}, namely:
\be
d_{Q} T(x_{1},\dots,x_{n}) = i \sum_{l=1}^{n} 
{\partial \over \partial x^{\mu}_{l}} T^{\mu}_{l}(x_{1},\dots,x_{n})
\label{gauge-inf}
\ee
for some auxiliary chronological products
$T^{\mu}_{l}(x_{1},\dots,x_{n}), \quad l = 1,\dots,n$
which must be determined recurringly, together with the standard chronological
products, and to construct the $S$-matrix
$S(g)$
for a test function $g$, that is without performing the adiabatic limit
$g \searrow 1.$ 

However, this point of view is not without problems. Indeed, if one imposes
(\ref{gauge-inf}) instead of (\ref{gi-Tn}), then the $S$-matrix so
constructed will {\it not} factorize to the physical space
$Ker(Q)/Im(Q)$
which raises the question about its physical relevance. To this one must add
the rather unpleasant fact that one abandons the consistency condition
(\ref{gi-Tn}) which has a direct physical relevance (the possibility of
constructing an $S$-matrix in the physical space
$Ker(Q)/Im(Q)$)
for an independent postulate (\ref{gauge-inf}). On the other hand,
the rather close connection between (\ref{gi-Tn}) and
(\ref{gauge-inf}) suggests that there must exists a common ``cure"
for both types of problems. That is, if one can find a reasonable solution of
the adiabatic limit problem, then it is reasonable to conjecture that one will
be able to strengthen the mathematical status of (\ref{gi-Tn}) and, eventually,
prove its equivalence with (\ref{gauge-inf}). In this case the consistency
condition can be also written in the following form:
\be
\left. d_{Q} \int_{(\R^{4})^{\times n}} dx_{1} \cdots dx_{n}
g_{\epsilon}(x_{1}) \cdots g_{\epsilon}(x_{n}) 
T(x_{1},\dots,x_{n}) \right|_{Ker(Q)} = {\cal O}(\epsilon), 
\quad \forall n \geq 1
\label{gi-Tn-epsilon}
\ee
in the sense of infinitesimal calculus of Dieudonn\'e. In what follows, the
interpretation of the right hand side of the preceding relations will be ``a
integrated divergence". In other words, to avoid various problems we will use
in fact the {\it formal adiabatic limit condition} given by (\ref{gauge-inf}).
A more detailed discussion on this point can be found in \cite{Gr2}. 

By a {\it trivial Lagrangian} we mean a Wick expression of the type
\be
L(x) = d_{Q} N(x) + i {\partial \over \partial x^{\mu}} L^{\mu}(x)
\label{T1-trivial}
\ee
with
$L(x)$
and
$L^{\mu}(x)$
some Wick polynomials. The first term in the previous formula gives zero
by factorisation to the physical Hilbert space (according to a previous
discussion) and the second one gives also zero in the adiabatic limit; this
justify the elimination of such expression from the first order chronological
product
$T(x)$.

If one completely exploits the condition of gauge invariance in the first order
of perturbation theory obtaining the generic form of the Yang-Mills interaction
of spin-one Bosons up to a trivial Lagrangian. We assume the summation 
convention of the dummy indices
$
a,b,\dots = 1,\dots,r.
$
The result from \cite{Gr2} is:
                        
\begin{thm}
Let us consider the operator
$T(x)$
defined on 
${\cal H}^{gh,r}_{YM}$
as a Lorentz-invariant Wick polynomial in 
$
A^{\mu}_{a}(x),~u_{a}(x),~\tilde{u}_{a}(x),~\Phi_{a}(x)
$
such that every term has canonical dimension 3 of 4. If it verifies the formal
adiabatic limit condition then it has, up to a trivial Lagrangian, the
following form:
\bea
T^{YM}(x) = 
f_{abc} \left[ :A_{a\mu}(x)A_{b\nu}(x) \partial^{\nu} A_{c}^{\mu}(x):
- :A_{a}^{\mu}(x) u_{b}(x) \partial_{\mu} \tilde{u}_{c}(x):\right],
\nonumber \\
+ f'_{abc} \left[ :\Phi_{a}(x) \partial_{\mu} \Phi_{b}(x) A_{c}^{\mu}(x): 
- m_{b} :\Phi_{a}(x) A_{b\mu}(x) A_{c}^{\mu}(x): 
- m_{b} :\Phi_{a}(x) \tilde{u}_{b}(x) u_{c}(x):\right]
\nonumber \\
+ f^{"}_{abc} :\Phi_{a}(x) \Phi_{b}(x) \Phi_{c}(x): 
+ g_{abcd} :\Phi_{a}(x) \Phi_{b}(x) \Phi_{c}(x) \Phi_{d}(x): 
\label{YM-1}
\eea
                        
The various constants from the preceding expression are constrained by the
following conditions:
                        
- the expressions 
$f_{abc}$
are completely antisymmetric
\be
f_{abc} = - f_{bac} = - f_{acb}
\label{anti-f}
\ee
and verify:
\be
(m_{a} - m_{b}) f_{abc} = 0, \quad {\rm iff} \quad m_{c} = 0, \quad 
\forall a,b = 1,\dots,r;
\label{mass-f}
\ee
                        
- the expressions
$f'_{abc}$
are antisymmetric  in the indices $a$ and $b$:
\be
f'_{abc} = - f'_{bac},
\label{anti-f'}
\ee
verify the relation:
\be
(m^{H}_{a} - m^{H}_{b}) f'_{abc} = 0, \quad 
{\rm iff} \quad m_{a} = m_{b} = m_{c} = 0,
\quad \forall a,b = 1,\dots,r
\label{mass-f'}
\ee
and are connected to 
$f_{abc}$
by:
\be
f_{abc} m_{c} = f'_{cab} m_{a} - f'_{cba} m_{b}, \quad 
\forall a,b,c = 1,\dots,r;
\label{f-f'}
\ee
                        
- the expressions 
$f^{"}_{abc}$
remain undetermined for
$m_{a} = m_{b} = m_{c} = 0$
and for the opposite case are given by:
\be
f^{"}_{abc} = 
{1 \over 6m_{c}} f'_{abc} 
\left[(m^{*}_{a})^{2} - (m^{*}_{b})^{2} - m_{a}^{2} + m_{b}^{2}\right],
\label{f"}
\ee
for
$m_{c} \not= 0$.

- the expressions
$g_{abcd}$
are non-zero only for
$m_{a} = m_{b} = m_{c} = m_{d} = 0$
and in this case they are completely symmetric. 
\label{T1}
\end{thm}

\begin{rem}
The presence of indices of type IIb and IIc is taken into account by requiring
that the constants from 
$T(x)$
are null if one of the indices 
$a, b, c$
takes such values. One can see that this does not affect the equations from the
statement of the theorem.
\end{rem}
We also have:      
\begin{cor}
In the condition of the preceding theorem, one has:
\be
d_{Q} T(x) = i\partial_{\mu} T^{\mu}(x)
\label{k1}
\ee
where:
\be
T^{\mu} = f_{abc} \left( :u_{a} A_{b\nu} F^{\nu\mu}_{c}: -
{1\over 2} :u_{a} u_{b} \partial^{\mu} \tilde{u}_{c}: \right)
+ f'_{abc} \left( m_{a} :A_{a}^{\mu} \Phi_{b} u_{c}:
+ :\Phi_{a} \partial^{\mu}\Phi_{b} u_{c}: \right).
\label{Tmu}
\ee
\end{cor}
                        
The expression
$T(x)$
from the preceding theorem verifies the unitarity condition
$$
T(x)^{\dagger} = T(x)
$$
if and only if the constants 
$
f_{abc},~ f'_{abc}
$
and
$
f"_{abc},
$
have real values; it also verifies the causality condition:
$$
[T(x), T(y)] = 0,\quad \forall x, y \in \R^{4} \quad {\rm s.t.} \quad
(x-y)^{2} < 0. 
$$

We close this Subsection with some remarks.

\begin{rem}
One can see that the necessity of using ghost fields stems from the fact that
there seems to be impossible to construct the interaction Lagrangian without
them. However, from a fundamental point of view, one can consider them only as
some catalysers \cite{DS1} and hope that one will be able to reformulate the
whole theory without them.
\end{rem}
\begin{rem}
In the first order analysis one can also use instead of the formal adiabatic
limit condition (\ref{gauge-inf}) the more physical condition (\ref{gi-Tn})
because no problems connected with the adiabatic limit exists in this case.
However, as notices in \cite{DS}, the condition does essentially eliminate the
tri-linear terms and one looses a lot of the information of the preceding
theorem. This is another indication that one should work with the formal
adiabatic limit condition.
\end{rem}
\begin{rem}
In \cite{Du1} one can find a discussion showing that trivial Lagrangians do not
produce effects in the higher orders of perturbation theory.
\end{rem}

\subsection{Yang-Mills Fields coupled to Matter}
                        
We study here the possibility of coupling Yang-Mills fields to ``matter". We 
suppose that we are given the Hilbert space of ``matter" 
${\cal H}_{matter}$
which is ussually also a Fock space. Then the coupled system is described in
the tensor product Hilbert space
${\cal F}_{YM} \otimes {\cal H}_{matter}$.
One can describe this Fock space considering 
$
\tilde{\cal H}^{gh,r}_{YM} \equiv 
{\cal H}^{gh,r}_{YM} \otimes {\cal H}_{matter}
$
with the corresponding supercharge operator and forming the quotient
$Ker(Q)/Im(Q)$. We will consider here that the ``matter" is formed from Dirac
Fermions only. 
                        
First, we generalize theorem \ref{T1}:
\begin{thm}
Let us consider the operator
$T(x)$
defined on 
$\tilde{\cal H}^{gh,r}_{YM}$
which is a Lorentz-invariant Wick polynomial in 
$A_{a}^{\mu}(x),~u_{a}(x)~,\tilde{u}_{a}(x)~,\Phi_{a}(x)$
and the matter fields such that every term has canonical dimension 3 or 4. 
Then 
$T(x)$
verifies the formal adiabatic limit condition, if and only if, up to a trivial
Lagrangian it has the following form:
\bea
T(x) = T^{YM}(x) + A_{a}^{\mu}(x) j_{a\mu}(x) +
\sum_{m_{a} \not= 0} {1 \over m_{a}} \Phi_{a}(x) \partial_{\mu}j^{\mu}_{a}(x)
%\nonumber \\
+ \sum_{m_{a} = 0} \Phi_{a}(x) j_{a}(x) + T_{\rm matter}(x)
\label{YM-1-matter}
\eea
Here 
$T^{YM}(x)$
has been defined in theorem \ref{T1},
$j_{a\mu}$
and
$j_{a}$
are Lorentz covariant currents build only from the matter fields with
$\omega(j_{a\mu}) = 1,2,3$
and
$T_{matter}(x)$
contains only the matter fields. Moreover the following conservation law should
be valid:
\be
\partial_{\mu} j^{\mu}_{a}(x) = 0, \quad \forall m_{a} = 0.
\label{conservation}
\ee
                        
The expression for 
$T(x)$
verifies the unitarity requirement if and only if we have:
\be
j_{a}^{\mu}(x)^{\dagger} = j_{a}^{\mu}(x),\quad \forall a = 1,\dots,r, \quad
j_{a}(x)^{\dagger} = j_{a}(x), \quad \forall m_{a} = 0
\ee
and verifies the causality condition if and only if:
\be
[j_{a}^{\mu}(x), j_{b}^{\nu}(y) ] = 0,\quad (x-y)^{2} < 0, \quad
\forall a, b = 1,\dots,r,
\ee
\be
[j_{a}(x), j_{b}(y) ] = 0,\quad (x-y)^{2} < 0, \quad \forall m_{a} = m_{b} = 0,
\ee
\be
[j_{a}^{\mu}(x), j_{b}(x) ] = 0,\quad (x-y)^{2} < 0, \quad \forall m_{b} = 0.
\label{comm-j}
\ee
\label{T1-matter}
\end{thm}
                        
{\bf Proof:}
                  
Beside the terms considered in theorem \ref{T1} we have to include terms
containing explicitly the Dirac Fermions. Lorentz covariance and power counting
limit these terms to
$T_{\rm matter}(x)$
and:
\be
T_{\rm matter}(x) \equiv A^{\mu}_{a}(x) j_{a\mu}(x) + \Phi_{a}(x) j_{a}(x)
\label{dirac}
\ee
with 
$j_{a\mu}~(j_{a})$
a Lorentz covariant (resp. invariant) operator. Proceeding in the same way as
for the proof of theorem \ref{T1}, we obtain a supplementary restriction,
namely:
\be
m_{a} j_{a} = \partial_{\mu} j^{\mu}_{a}, \quad \forall a = 1,\dots,r.
\label{gau}
\ee
                        
In other words, for 
$m_{a} = 0$
we get (\ref{conservation}) and for
$m_{a} \not= 0$
we get:
\be
j_{a} = {1\over m_{a}} \partial_{\mu} j^{\mu}_{a}.
\ee
                        
The expression from the statement emerges. The other assertions are
straightforward, although rather tedious to verify.
$\qed$

It is clear if the Hilbert space of the matter fields is also a Fock space and
the currents are build from Wick monomials, then the commutation relations
(\ref{comm-j}) are always verified.
                        
\begin{cor}
The following formula is true
\be
d_{Q} T(x) = i {\partial \over \partial x^{\mu}} T^{\mu}(x)
\label{dQT1}
\ee
where
$
T^{\mu}(x)
$
is obtained by adding to the corresponding expression from the pure Yang-Mills
case - see (\ref{Tmu}) - the following contribution due to the presence of
matter:
\be
T^{\mu}_{\rm matter}(x) \equiv u_{a}(x) j^{\mu}_{a}(x).
\label{Tmu-mat}
\ee
\end{cor}
                        
Now we get in detail the structure of the interaction Lagrangian in the
following two propositions. We have:
                        
\begin{prop}
Suppose that the Dirac Fermions generating
${\cal H}_{matter}$
are
$\psi_{A}$
of masses
$M_{A} \geq 0, \quad A = 1,\dots,N$.
Then the generic form of the currents from the preceding theorem are:
\be
j_{a}^{\mu}(x) = 
:\overline{\psi_{A}}(x) (t_{a})_{AB} \gamma^{\mu} \psi_{B}(x): +
:\overline{\psi_{A}}(x) (t'_{a})_{AB} \gamma^{\mu} \gamma_{5} \psi_{B}(x):
\label{vector-current}
\ee
and
\be
j_{a}(x) = 
:\overline{\psi_{A}}(x) (s_{a})_{AB} \psi_{B}(x): +
:\overline{\psi_{A}}(x) (s'_{a})_{AB} \gamma_{5} \psi_{B}(x):
\label{scalar-current}
\ee
                        
The causality conditions from theorem \ref{T1-matter} are fulfilled and the
hermiticity conditions are equivalent with the fact that the complex
$N \times N$
matrices
$t_{a},\quad t'_{a}, \quad s_{a}, \quad a = 1,\dots r$
are hermitian and
$s_{a}', \quad a = 1,\dots,r$
anti-hermitian. 
\end{prop}
                        
The contributions with (without) the matrix
$\gamma_{5}$
is called {\it axial} (resp. {\it vectorial}) current. Let us define the 
{\it mass matrix} by:
\be
M_{AB} \equiv \delta_{A,B} M_{A}, \quad \forall A, B = 1,\dots,N.
\ee
                        
Then we have:
\begin{prop}
The following mass relations are true:
\be
s_{a} = {i \over m_{a}} [M, t_{a}], \quad
s_{a}' = {i \over m_{a}} \{M, t'_{a}\}, \quad \forall m_{a} \not= 0,
\label{mass1}
\ee
\be
[M, t_{a}] = 0, \quad  \{M, t'_{a}\} = 0, \quad \forall m_{a} = 0.
\label{mass2}
\ee
                        
In particular, the matrices 
$t_{a}, \quad \forall m_{a} = 0$
can be exhibited into a block diagonal structure (eventually after a
relabelling of the Dirac fields) and the masses corresponding to the same block
must be equal. 
\end{prop}
                        
{\bf Proof:}
It is easy to show that the conservation law ({\ref{gau}) is equivalent to the
two relations from the statement.  
$\qed$

\begin{cor}
Let us define
\be
t_{a}^{\epsilon} \equiv t_{a} + \epsilon t'_{a}, \quad
s_{a}^{\epsilon} \equiv s_{a} + \epsilon s'_{a}, \quad
\label{epsi}
\forall a = 1,\dots,r, \quad \epsilon = \pm;
\ee
here
$\epsilon = \pm$.
Then, the relations (\ref{mass1}) and (\ref{mass2}) are equivalent to:
\be
s_{a}^{\epsilon} = {i \over m_{a}} (M t_{a}^{\epsilon} - t_{a}^{-\epsilon} M),
\quad \forall m_{a} \not= 0,
\label{s}
\ee
\be
M t_{a}^{\epsilon} = t_{a}^{-\epsilon} M, \quad \forall m_{a} = 0
\label{s0}
\ee
and the hermiticity conditions are equivalent to:
\be
(t_{a}^{\epsilon})^{*} = t_{a}^{\epsilon}, \quad 
(s_{a}^{\epsilon})^{*} = s_{a}^{-\epsilon}, \quad \forall a = 1,\dots,r, \quad
\epsilon = \pm.
\ee
\end{cor}

\newpage
\section{Perturbation Theory\label{perturbation}}

\subsection{The General Framework\label{bogoliubov}}

We give here the basic ideas  of a {\it multi-Lagrangian} perturbation
theory following \cite{EG1} and \cite{Gr3}. One can argue that the 
$S$-matrix is formal series of operator valued distributions:
\be
S({\bf g})=1 + \sum_{n=1}^\infty{i^{n}\over n!}\int_{\R^{4n}} 
dx_{1}\cdots dx_{n}\, T_{j_{1},\dots,j_{n}}(x_{1},\cdots, x_{n})
g_{j_{1}}(x_{1})\cdots g_{j_{n}}(x_{n}),
\label{S}
\ee
where
${\bf g} = \left( g_{j}(x)\right)_{j = 1, \dots P}$
is a multi-valued tempered test function in the Minkowski space 
$\R^{4}$
that switches the interaction and
$T_{j_{1},\dots,j_{n}}(x_{1},\cdots, x_{n})$
are operator-valued distributions acting in the Fock space of some collection
of free fields with a common dense domain of definition
$D_{0}$. 
These operator-valued distributions are called {\it chronological products} and
verify some properties called {\it Bogoliubov axioms}. We note that there is a
canonical association of the point $x_{i}$ and the index $j_{i}$.  One starts
from a set of {\it interaction Lagrangians}
$T_{j}(x), \quad j = 1,\dots, P$
and tries to construct the whole series
$T_{j_{1},\dots,j_{n}}, \quad n \geq 2$.

We outline briefly the set of axioms imposed on the chronological products 
$T_{j_{1},\dots,j_{n}}$;
we do not give the explicit formul\ae~ because they are well known in the
literature and can be found in the references quoted above. 
\begin{itemize}
\item
Symmetry: this axiom describes the behaviour of the chronological products with
respect to the permutation of the couples 
$(x_{i},j_{i})$.
\item
Poincar\'e invariance: this axiom describes the behaviour of the chronological
products with respect to the action of the Poincar\'e group in the Fock space
of the system. Essentially is a tensorial covariance condition.

\item
Causality: it describes factorization properties of the chronological products
for causally separated arguments. This seems to be the central axiom of this
axiomatic approach; it plays a major r\^ole in other axiomatic schemes as well.

\item 
Unitarity: this axiom is considered in the sense of formal series.
\end{itemize}

A {\it renormalization theory} is the possibility to construct such a
$S$-matrix starting from the first order terms:
$
T_{j}(x), \quad j = 1,\dots, P
$
which are linearly independent Wick polynomials called 
{\it interaction Lagrangians} which should verify the corresponding axioms
expressing the behaviour with respect to Poincar\'e transformations, Hermitian
conjugation and commutation properties for space-like separated arguments.

The case of a single Lagrangian corresponds to a single coupling constant, that
is
$P = 1$
and in that case the chronological products will be operators
$T(X)$
without any indices. However, it is more convenient to consider that the
interaction Lagrangian is given by the sum
\be
T(x) = \sum c_{j} T_{j}(x)
\label{one}
\ee
with 
$c_{j}$
some real constants. In this case, the chronological products of the theory are
\be
T(X) = \sum c_{j_{1}} \dots c_{j_{n}} T_{j_{1},\dots,j_{n}}(X).
\label{one-n}
\ee

It can be showed that that one must consider the given interaction 
Lagrangians 
$T_{j}(x)$
to be all Wick monomials canonical dimension 
$\omega_{j} \leq 4$
($j = 1,\dots,P$)
acting in the Fock space of the system.  Because the Fock space is generated by
some free relativistic fields acting on the vacuum
$\Omega$
it is easy to see that there always have covariance properties with respect to
Poincar\'e transformations. 

If there are non-Hermitian free fields acting in the Fock space, we have in 
general:
\be
T_{j}(x)^{\dagger} = T_{j^{*}}(x)
\label{skew-unitarity1}
\ee
where 
$j \rightarrow j^{*}$
is a bijective map of the numbers
$1,2,\dots,P$.

If there are Fermi or ghost fields acting in the Fock space, the causality
property is in general:
\be
T_{j_{1}}(x_{1}) T_{j_{2}}(x_{2}) = 
(-1)^{\sigma_{j_{1}}\sigma_{j_{2}}} T_{j_{2}}(x_{2}) T_{j_{1}}(x_{1}),
\quad \forall x_{1} \sim x_{2}.
\label{skew-causality1}
\ee

Here
$\sigma_{i}$
is the number of Fermi and ghost fields factors in the Wick monomial
$T_{j}$;
if
$\sigma_{j}$
is even (odd) we call the index
$j$
{\it even} (resp. {\it odd}). One has to keep track of these signs in the
symmetry axiom for the chronological products.

It is convenient to let the index $j$ have the value $0$ also and we
put by definition
\be
T_{0} \equiv {\bf 1}.
\ee

Moreover, we define a new sum operation of two indices 
$j_{1}, j_{2} = 1,\dots,P$;
this summation is denoted by $+$ but should not be confused with the ordinary
sum. By definition we have:
\be
T_{j_{1}+j_{2}}(x) = c :T_{j_{1}}(x) T_{j_{2}}(x):
\ee
for some positive constant $c$. We define componentwise the summation for 
$n$-tuples
$J = \{j_{1},\dots,j_{n}\}$. 
The new summation is non-commutative if Fermi or ghost fields are present.

We will use the notation
\be
\omega_{J} \equiv \sum_{j \in J} \omega_{j}
\ee
and we call it the {\it canonical dimension} of 
$T_{J}(X)$.

Let us denote by
$\omega(d)$
the order of singularity of the numerical distribution $d$. we use the
definition from \cite{Sc1} although one can also use the scaling degree
introduced by Steinmann (see \cite{DF}).

Then we add a new axiom, namely the following {\it Wick expansion property}
of the chronological products is valid:
\be
T_{J}(X) = \sum_{K+L=J} \epsilon \quad t_{K}(X) \quad W_{L}(X)
\label{wick-chrono}
\ee
where: (a)
$t_{K}(X)$
are numerical distributions (the {\it renormalized Feynman amplitudes});
(b) the degree of singularity is restricted by the following relation:
\be
\omega(t_{K}) \leq \omega_{K} - 4(n-1);
\label{deg-chrono}
\ee
(c) $\epsilon$ is the sign coming from permutation of Fermi fields;
(d) we have introduced the notation
\be
W_{J}(X) \equiv :T_{j_{1}}(x_{1})\cdots T_{j_{n}}(x_{n}):
\ee

Let us notice that from (\ref{wick-chrono}) we have:
\be
t_{J}(X) = < \Omega, T_{J}(X) \Omega >.
\label{average-chrono}
\ee 

In particular, these numerical distributions have causal support and are
Poincar\'e covariant; translation invariance implies that they are in fact
distributions in
$m = 4 (|X|-1)$
variables.

The recursive construction assumes that we have the expressions
$T_{J}(X)$
for
$|X| \leq n - 1$
verifying all the properties and tries to construct them for
$X = n$.
The basic object is the commutator function:
\be
D_{j_{1},\dots,j_{n}}(x_{1},\dots,x_{n-1};x_{n}) \equiv 
A'_{j_{1},\dots,j_{n}}(x_{1},\dots,x_{n-1};x_{n}) -
R'_{j_{1},\dots,j_{n}}(x_{1},\dots,x_{n-1};x_{n})
\label{com-D}
\ee
where
\be
A'_{j_{1},\dots,j_{n}}(x_{1},\dots,x_{n-1};x_{n}) \equiv 
{\sum}'_{X_{1},X_{2} \in Part(X)}
(-1)^{|X_{2}|} T_{J_{1}}(X_{1}) \bar{T}_{J_{2}}(X_{2})
\ee
and
\be
R'_{j_{1},\dots,j_{n}}(x_{1},\dots,x_{n-1};x_{n}) \equiv 
{\sum}'_{X_{1},X_{2} \in Part(X)}
(-1)^{|X_{2}|} \bar{T}_{J_{2}}(X_{2}) T_{J_{1}}(X_{1});
\ee
the sums $\sum'$ run over the partitions verifying
$X_{2} \not= \emptyset, \quad x_{n} \in X_{1}$.

The commutator function can be proved to be Poincar\'e covariant and
to have causal support i.e.
$supp(D_{j_{1},\dots,j_{n}}(x_{1},\dots,x_{n-1};x_{n})) 
\subset \Gamma^{+}(x_{n}) \cup \Gamma^{-}(x_{n})$
where we use standard notations:
\be
\Gamma^{\pm}(x_{n}) \equiv \{ (x_{1},\dots,x_{n}) \in (\R^{4})^{n} |
x_{i} - x_{n} \in V^{\pm} , \quad \forall i = 1, \dots, n-1\}.
\ee

Moreover, a formula similar to (\ref{wick-chrono}) is true:
\be
D_{J}(X) = \sum_{K+L=J} \epsilon \quad d_{K}(X) \quad W_{L}(X)
\label{wick-d}
\ee
where
$d_{K}(X)$
are numerical distributions; in analogy to (\ref{average-chrono}) we have:
\be
d_{J}(X) = < \Omega, D_{J}(X) \Omega >.
\label{average-d}
\ee 

It follows that the numerical distributions
$d_{J}(X)$
have causal support i.e
$
supp(d_{J}(X)) \subset \Gamma^{+}(x_{n}) \cup \Gamma^{-}(x_{n})
$
and are 
$SL(2,\C)$-invariant. Moreover, their degree of singularity is restricted by
\be
\omega(d_{K}) \leq \omega_{K} - 4(n-1);
\label{deg-d}
\ee
this is the content of the {\it power counting theorem}. One knows that there
exists a causal splitting
\be
d_{J} = a_{J} - r_{J}, \quad supp(a_{J}) \subset \Gamma^{+}(x_{n}), \quad 
supp(r_{J}) \subset \Gamma^{-}(x_{n})
\label{causal-splitting}
\ee
which is also
$SL(2,\C)$-invariant and such that the order of the singularity is preserved.
So, there exists a 
$SL(2,\C)$-covariant causal splitting:
\be
D_{J}(X) = A_{J}(X) -  R_{J}(X), \quad |X| = n
\label{decD}
\ee
with
$supp(A_{j_{1},\dots,j_{n}}(x_{1},\dots,x_{n-1};x_{n})) \subset 
\Gamma^{+}(x_{n})$
and
$supp(R_{j_{1},\dots,j_{n}}(x_{1},\dots,x_{n-1};x_{n})) \subset 
\Gamma^{-}(x_{n})$.

Let us define
\be
T_{J}(X) \equiv A_{J}(X) - A_{J}'(X) = R_{J}(X) - R_{J}'(X).
\label{chronos-n}
\ee
Then these expressions satisfy the 
$SL(2,\C)$-covariance, 
and causality axioms. One can also fix unitarity and symmetry. 

We end this Subsection with an important remark. Let us consider some general
Wick polynomials
\be
A_{i}(x) = \sum_{j} c_{ij} \quad T_{j}(x), \quad i = 1,2,\dots
\ee

Then we can define the chronological products:
\be
T(A_{1}(x_{1}),\cdots,A_{n}(x_{n})) \equiv \sum_{J} 
c_{i_{1}j_{1}} \cdots c_{i_{n}j_{n}} \quad
T_{j_{1},\dots,j_{n}}(x_{1},\cdots, x_{n}).
\ee

One can find in \cite{DF} a system of axioms for the expressions
$T(A_{1}(x_{1}),\cdots,A_{n}(x_{n}))$
which is equivalent to the Bogoliubov set of axioms.

%\newpage
\subsection{Ward Identities\label{st}}

As we have said in the Subsection \ref{ym} the problem is to construct
the whole series 
$T(X)$
such that one has the gauge invariance condition in all orders of the
perturbation theory at the same time with the other Bogoliubov axioms.

In general we have something more general than relation (\ref{one}):
\be
T(x) = \sum c_{j} T_{j}(x)
\quad
T^{\mu}(x) = \sum c^{\mu}_{j} T_{j}(x)
\label{t1}
\ee
with 
$c_{j}, \quad c^{\mu}_{j}$
some real constants; then we will have something more general than 
(\ref{one-n}):
\be
T(X) = \sum c_{j_{1}} \cdots c_{j_{n}} T_{j_{1},\dots,j_{n}}(X),
\quad
T^{\mu}_{l}(X) = \sum c_{j_{1}} \dots c^{\mu}_{j_{l}} \dots c_{j_{n}} 
T_{j_{1},\dots,j_{n}}(X)
\label{tn}
\ee

In particular, the following conventions hold:
\be
T(\emptyset) \equiv {\bf 1}, \quad 
T^{\mu}_{l} (\emptyset) \equiv 0, \quad
T^{\mu}_{l} (X) \equiv 0, \quad {\rm for} \quad x_{l} \not\in X.
\label{empty-p}
\ee

Then the gauge invariance condition (\ref{gauge-inf}) can be written more
compactly as follows:
\be
d_{Q} T(X) = i \sum {\partial \over \partial x^{\mu}_{l}} T^{\mu}_{l}(X).
\label{gauge}
\ee

We suppose that these relations are true up to order 
$|X| \leq n - 1$ 
and investigate the possible obstructions in order $n$. The procedure used in
\cite{DHKS1}, \cite{DHKS2} and \cite{Gr1}, \cite{Gr2} amounts to the following.
Let us define the operator distributions 
$D(X)$ 
and
$D^{\mu}_{l}(X)$ 
in analogy to the relations (\ref{tn}). Then it can be proved that we have:
\be
d_{Q} D(X) = i \sum_{l=1}^{n} {\partial \over \partial x^{\mu}_{l}}
D^{\mu}_{l}(X), \quad |X| = n.
\label{gauge-D}
\ee

We can express this condition in terms of numerical distributions. According to
the relation (\ref{wick-chrono}) and Wick theorem we must have Wick expansions
for the two expressions appearing in the preceding equation:
\be
D(X) = \sum_{J} d_{J}(X) W_{J}(X),
\quad
D^{\mu}_{l}(X) = \sum_{J} d^{\mu}_{l;J}(X) W_{J}(X)
\label{wick-dd}
\ee

The numerical distributions appearing in these relations have the following
properties: they are Poincar\'e covariant, they have causal support and the
order of singularity can be restricted according to the {\it power counting}
formula:
\be
\omega(d_{J}) + \omega_{J} \leq 4,
\quad
\omega(d^{\mu}_{l;J}) + \omega_{J} \leq 4
\label{deg-feynman}
\ee
according to the power counting theorem.

One can rewrite (\ref{wick-dd}) as follows:
\be
D(X) = \sum_{i} d_{i}(X) W_{i}(X),
\quad
D^{\mu}_{l}(X) = \sum_{i} d_{i}(X) W^{\mu}_{l;i}(X)
+ \sum_{i} d^{\mu}_{i}(X) W_{l;i}(X)
\label{wick-di}
\ee
where
$d_{i}$
and
$d_{i}^{\mu}$
can be taken linear independent over the vector space of smooth functions with
polynomial bounded increase at infinity
${\cal O}_{M}$.
The index $i$ takes a finite number of values and the expressions
$W_{i}(X),~ W_{l;i}(X),~ W^{\mu}_{l;i}(X)$
are Wick polynomials.

Using the linear independence one obtains from (\ref{gauge-D}) a set of
identities among Wick polynomials of the type
\be
d_{Q} W_{i} = \cdots
\label{gauge-w}
\ee
where the left hand side can be computed as follows. First one makes the
derivation operations in the right hand side of (\ref{gauge-D}). It is quite
possible that relations of the type
\be
{\partial \over \partial x^{\mu}_{l}} d^{\mu}_{l;i}(X) = 
\sum_{j} c_{j} d_{j}(X)
\label{ward}
\ee
are valid for some numbers
$c_{j}$.
Then one has to rearrange the expression in the right hand side of
(\ref{gauge-D}) and the right hand side of (\ref{gauge-w}) emerges as the
coefficient of
$d_{i}(X)$.

Identities of the type (\ref{ward}) are called {\it Ward-Takahashi} (or {\it
Slavnov-Taylor identities}). In \cite{DHKS2} these relations are called the
{\it C-g identities}. They have been extensively studied in \cite{Du2}. In
lower orders of perturbation theory one can check them by explicit computation.

One now can interpret the renormalization theory as a distribution-splitting
preserving of the Ward identities. Suppose that one can find a causal splitting
$d_{i} = d^{adv}_{i} - d^{ret}_{i}$
of the set of causal distributions
$d_{i}(X)$
such that we preserve Poincar\'e covariance, the order of singularity and the
identities (\ref{ward}) i.e. we also have:
\be
{\partial \over \partial x^{\mu}_{l}} (d^{\mu}_{l;i})^{adv(ret)}(X) = 
\sum_{j} c_{j} d^{adv(ret)}_{j}(X)
\label{ward-a}
\ee

Then we define the expressions
$A(X)$
and
$A^{\mu}_{l}(X)$
by making into the formul\ae~ (\ref{wick-di}) the substitutions 
$d \rightarrow d^{adv}$.
If we use now the relations (\ref{gauge-w}) we easily obtain
\be
d_{Q} A(X) = i \sum_{l=1}^{n} {\partial \over \partial x^{\mu}_{l}}
A^{\mu}_{l}(X), \quad |X| = n.
\label{gauge-A}
\ee

The similar property for the chronological products of order $n$ easily
follows. So, the obstructions to the gauge invariance in order $n$ can appear
in the process of causally splitting the relations (\ref{ward}) i.e. we might
have instead of (\ref{ward-a}):
\be
{\partial \over \partial x^{\mu}_{l}} (d^{\mu}_{l;i})^{adv(ret)}(X)
- \sum_{j} c_{j} d^{adv(ret)}_{j}(X) = p(X)
\label{ward-ano}
\ee
where the expression in the right hand side
$p(X)$
- called {\it anomaly} - must have the form
\be
p(X) = p(\partial) \delta(X);
\ee
here
$p(\partial)$
is a Lorentz covariant polynomial in the partial derivative operators and
\be
\delta(X) \equiv \delta(x_{1} - x_{n}) \cdots \delta(x_{n-1} - x_{n}).
\ee

Also, if the distribution appearing in (\ref{ward}) have some global symmetry
property (symmetry with respect to some global group of symmetries,
(anti)symmetry with respect to some indices, etc.) one can usually perform the
distribution splitting such that these properties are preserved also.
Moreover, we have a limitation on the degree of the polynomial
$p(\partial)$:
\be
deg(p) \leq \omega
\ee
where $\omega$ is the order of singularity of the left hand side of
(\ref{ward-ano}). It easily follows a case when there are no anomalies, namely
when:
$\omega(d^{\mu}_{i}) \leq -2, \quad \forall \mu$.
Let us note in closing this Section that the form of a anomaly can be
simplified by redefinitions of the distributions
$a_{i}$
and
$a^{\mu}_{l;i}$;
we have the freedom of adding expressions of the type
$p(\partial) \delta(X)$.

\newpage
\section{Second Order Perturbation Theory\label{second}}        
 
\subsection{Yang-Mills coupled to Matter}

We follow \cite{Gr2} where the pure Yang-Mills case was studied emphasizing the
possible appearance of anomalies in a more explicit way. 
                        
\begin{thm}
Suppose that the distribution
$T(x,y)$
verifies (\ref{deg-chrono}). The it verifies the formal adiabatic limit
condition if and only if the following identities are verified:
                        
\be
f_{abc}f_{dec} + f_{bdc} f_{aec} + f_{dac} f_{bec} = 0, 
\quad a, b, d, e = 1,\dots,r;
\label{Jacobi}
\ee
\be
f'_{dca} f'_{ceb} - f'_{dcb} f'_{cea} = - f_{abc} f'_{dec},
\quad a, b, d, e = 1,\dots,r;
\label{repr-f'}
\ee
\be
f'_{cab} f^{"}_{cde} + f'_{cdb} f^{"}_{cae} + f'_{ceb} f^{"}_{cda} = 0,
\quad {\rm iff} \quad m_{a} = m_{b} = m_{d} = m_{e} = 0;
\label{h3-0}
\ee
\be
{\cal S}_{bcdef}f'_{cba} g_{cdef} = 0,
\quad a, b, d, e, f = 1,\dots,r;
\label{f'g}
\ee
\be
[t_{a}^{\epsilon},t_{b}^{\epsilon}] = i f_{abc} t_{c}^{\epsilon}, 
\quad \epsilon = \pm, \quad a, b = 1,\dots,r;
\label{repr}
\ee
\be
t_{a}^{-} s_{b}^{+} - s_{b}^{+} t_{a}^{+} = i f'_{bca} s_{c}^{+},
\quad a, b = 1,\dots,r.
\label{WE}
\ee 
\label{T2}

Here
${\cal S}_{\dots}$
is the symmetrization operator in the indices which are explicitly exhibited.
\end{thm}
                        
{\bf Proof:}
(i) According to the ideas from Subsection \ref{st}, we compute the commutator
\be
D(x_{1},x_{2}) \equiv [T(x_{1}), T(x_{2})]
\label{D1}
\ee
using Wick theorem and identify a set of linearly independent distributions
$d_{i}$
as in (\ref{wick-di}); these are distributions in one variable 
$\xi \equiv x_{1} - x_{2}$
due to translation invariance. Direct inspection of the expressions
(\ref{YM-1}) and (\ref{YM-1-matter}) produces the a list of such distributions
$\Delta$ 
with causal support. Using Feynman graph terminology we have distributions
associated to tree and one, two and three loops graphs. All these distributions
can be written as sum of the positive (negative) frequency parts:
\be
\Delta = \Delta^{(+)} + \Delta^{(-)}.
\label{conv1}
\ee

(a) From tree graphs: 
\bea
D_{m}, \quad \partial_{\rho} D_{m}, \quad 
\partial_{\rho}\partial_{\sigma} D_{m},
\nonumber \\
S_{M}(x) \equiv (i\gamma\cdot \partial + M) D_{M}(x),
\label{distr0}
\eea
where
$D_{m}$
is the Pauli-Villars commutator distribution of causal support corresponding to
mass $m$ (see \cite{Gr2} for the definition) and 
$S_{M}$
is the similar distribution for a Dirac field of mass $M$. 

(b) From one-loop graphs we get new distributions with causal support. 

\bea
D^{(\pm)}_{m_{1},m_{2}} \equiv \pm D^{(\pm)}_{m_{1}}(x) D^{(\pm)}_{m_{2}}(x),
\nonumber \\
D^{(\pm)}_{m_{1},m_{2};\rho} \equiv \pm
D^{(\pm)}_{m_{1}} \partial_{\rho}D^{(\pm)}_{m_{2}} - (1 \leftrightarrow 2).
\nonumber \\
\partial_{\rho}D_{m_{1},m_{2}},
\nonumber \\
D^{(\pm)}_{m_{1},m_{2};\rho\sigma} \equiv \pm \left[
\partial_{\rho}D^{(\pm)}_{m_{1}} \partial_{\sigma}D^{(\pm)}_{m_{2}}
- D^{(\pm)}_{m_{1}} \partial_{\rho}\partial_{\sigma}D^{(\pm)}_{m_{2}} \right]
+ (1 \leftrightarrow 2)
\nonumber \\
P^{(\pm)}_{M_{1},M_{2}}(x) \equiv \pm 
Tr \left[ S_{M_{1}}^{(-)}(\mp x) S_{M_{2}}^{(+)}(\pm x)\right],
\nonumber \\
P^{(\pm)}_{M_{1},M_{2};\rho}(x) \equiv \pm Tr \left[ S_{M_{1}}^{(-)}(\mp x)
\gamma_{\rho} S_{M_{2}}^{(+)}(\pm x)\right],
\nonumber \\
P^{(\pm)}_{M_{1},M_{2},\rho\sigma}(x) \equiv \pm 
Tr \left[ \gamma_{\rho} S_{M_{1}}^{(-)}(\mp x)
\gamma_{\sigma} S_{M_{2}}^{(+)}(\pm x)\right],
\nonumber \\
\Sigma^{(\pm)}_{m,M} \equiv \pm D^{(\pm)}_{m} S_{M}^{(\pm)}.
\label{distr1}
\eea

We note that in the definition of
$D^{(\pm)}_{m_{1},m_{2};\rho}$
we have taken the antisymmetric part in the masses because the symmetric part
has been considered separately :it is the third distribution from the list.

(c) From two-loops graphs:
\bea
D^{(\pm)}_{m_{1},m_{2},m_{3}} \equiv 
D^{(\pm)}_{m_{1}} D^{(\pm)}_{m_{2}} D^{(\pm)}_{m_{3}},
\nonumber \\
\partial^{2}D_{m_{1},m_{2},m_{3}}
\nonumber \\
D^{(\pm)}_{m_{1},m_{2};m_{3}} \equiv 
\partial_{\mu}D^{(\pm)}_{m_{1}} \partial^{\mu}D^{(\pm)}_{m_{2}} 
D^{(\pm)}_{m_{3}},
\nonumber \\
P^{(\pm)}_{m;M_{1},M_{2}} \equiv \pm D^{(\pm)}_{m} P_{M_{1},M_{2}}^{(\pm)},
\nonumber \\
P^{(\pm)}_{m;M_{1},M_{2};\rho\sigma} \equiv \pm D^{(\pm)}_{m}
P_{M_{1},M_{2};\rho\sigma}^{(\pm)}
\label{distr2}
\eea

(d) From three-loops graphs:
\be
D^{(\pm)}_{m_{1},m_{2},m_{3},m_{4}}(x) \equiv \pm
D^{(\pm)}_{m_{1}} D^{(\pm)}_{m_{2}} D^{(\pm)}_{m_{3}} D^{(\pm)}_{m_{4}}
\label{distr3}
\ee

The distributions
$P^{\dots}_{\dots}$
are obtained from contractions of two vectorial currents. Let us note that one
also obtains distributions of the type
$Q^{\dots}_{\dots}$
from contractions of two axial currents. These distributions can be obtained 
directly from the corresponding distributions
$P^{\dots}_{\dots}$
by conveniently inserting two
$\gamma_{5}$
factors. However, the distributions of the type
$Q^{\dots}_{\dots}$
can be expressed in terms of
$P^{\dots}_{\dots}$
if one uses the identity
\be
\gamma_{5} S^{(\pm)}_{M} \gamma_{5} = - S^{(\pm)}_{-M}.
\label{5s5}
\ee

The distributions following from contractions of axial and a vectorial current 
are null because the traces so obtained are null.
Next, we note that in the other commutators
\be
D^{\mu}_{1}(x_{1},x_{2}) \equiv [T^{\mu}(x_{1}), T(x_{2})], \quad
D^{\mu}_{2}(x_{1},x_{2}) \equiv [T(x_{1}), T^{\mu}(x_{2})] =
- D^{\mu}_{1}(x_{2},x_{1})
\label{D1mu}
\ee
the distributions
$g_{\mu\lambda} d^{\lambda}_{l;i}$
from (\ref{wick-di}) can be of the following type:
\be
\partial_{\mu}D_{m}, \quad \gamma_{\mu} S_{M}, \quad D_{m_{1},m_{2};\mu}, \quad
D_{m_{1},m_{2};\mu\nu}, \quad P_{M_{1},M_{2};\mu}, \quad
P_{M_{1},M_{2};\mu\nu}
\ee
and the distributions of the type
$d_{i}$
can be of the type:
\be
D_{m}, \quad \partial_{\rho}D_{m}, \quad D_{m_{1},m_{2}}, \quad
D_{m_{1}m_{2};\rho}.
\ee

Here the various parameters 
$m, M, \dots$
are the masses appearing in the theory. If we consider distinct combinations of
masses and indices we obtain a linear independent set of distributions.

Let us also give for further use the orders of singularity of the distributions
listed above. We have:
\bea
\omega(D_{m}) = -2, \quad
\omega(D_{m_{1},m_{2}}) = 0, \quad
\omega(D_{m_{1},m_{2};\rho}) = -1, \quad
\omega(D_{m_{1},m_{2};\rho\sigma}) = 2, \quad
\omega(P_{M_{1},M_{2}}) = 2, \quad
\nonumber \\
\omega(P_{M_{1},M_{2};\rho}) = 1, \quad
\omega(P_{M_{1},M_{2};\rho\sigma}) = 2, \quad
\omega(P_{m;M_{1},M_{2}}) = 4, \quad
\omega(P_{m;M_{1},M_{2};\rho\sigma}) = 4,
\nonumber \\
\omega(\Sigma_{m,M}) = 1, \quad
\omega(D_{m_{1},m_{2},m_{3}}) = 2, \quad
\omega(D_{m_{1},m_{2};m_{3}}) = 4, \quad
\omega(D_{m_{1},m_{2},m_{3},m_{4}}) = 4. \qquad
\label{orders}
\eea

Some of these orders of singularity are in fact lower than naive power
counting suggests.

All these distributions have causal support so we have causal decompositions:
\be
\Delta = \Delta^{adv} - \Delta^{ret}
\label{conv2}
\ee

We have assumed that the causal splitting is preserving Lorentz covariance and
the order of singularity. If the order of singularity is less
$0$
then  this causal decomposition is a unique (see the end of the preceding
Section). This is the case for the distributions
$D_{m},~ S_{M}$
and
$D_{m_{1}m_{2};\rho}$.

(ii) Now we consider the Ward identities (\ref{ward}). By direct inspection one
finds out that they are:
\be
(\partial^{2} + m^{2} ) D_{m} = 0,
\label{G1}
\ee
\be
(i \gamma\cdot\partial - M) S_{M} = 
S_{M} (i\gamma\cdot\stackrel{\leftarrow}\partial - M) = 0,
\label{G2}
\ee
\be
\partial^{\mu}D_{m_{1},m_{2};\mu} = ( m_{2}^{2} - m_{1}^{2}) D_{m_{1},m_{2}},
\label{G3}
\ee
\be
\partial^{\mu}D_{m_{1},m_{2};\mu\nu} =
(m_{2}^{2} - m_{1}^{2}) D_{m_{1},m_{2};\nu}
\label{G4}
\ee
\be
\partial^{\mu}P_{M_{1},M_{2};\mu} = i (M_{1} - M_{2}) P_{M_{1},M_{2}}
\label{G5}
\ee
\be
\partial^{\mu}P_{M_{1},M_{2};\mu\nu} = i (M_{1} - M_{2}) P_{M_{1},M_{2};\nu}.
\label{G6}
\ee

Now we analyze possible anomalies resulting after the causal splitting
procedure. It is well known that the first two relations (\ref{G1}) and
(\ref{G2}) are indeed producing anomalies: for the (unique) causal splitting
considered above one gets:
\be
(\partial^{2} + m^{2} ) D^{adv(ret)}_{m} = \delta,
\label{G1-a}
\ee
\be
(i \gamma\cdot\partial - M) S^{adv(ret)}_{M} =
S^{adv(ret)}_{M} (i\gamma\cdot\stackrel{\leftarrow}\partial - M) = i\delta.
\label{G2-a}
\ee

One can prove more than that: even if we modify these splitting with arbitrary
local polynomial terms the anomalies do not disappear.

Next we consider (\ref{G4}); inspecting the orders of we can have the following
generic form of the anomaly:
\be
p_{\nu}(\partial) = c_{1} \partial_{\nu} + c_{3} \partial_{\nu} \partial^{2}.
\ee

We can eliminate this anomaly if we make the redefinition:
\be
D^{adv}_{m_{1},m_{2};\mu\nu} \rightarrow D^{adv}_{m_{1},m_{2};\mu\nu}
+ (c_{1} g_{\mu\nu} + c_{3} \partial_{\mu}\partial_{\nu} ) \delta.
\ee

The case (\ref{G6}) can be treated in a similar way and the anomaly is also
eliminated. The Ward identity (\ref{G3}) is non-trivial only for
$m_{1} \not= m_{2}$.
We have already noticed that there exists a unique causal decomposition
preserving Lorentz covariance and the order of singularity of
$
D_{m_{1},m_{2};\mu};
$
then we can {\bf define}:
\be
D^{adv}_{m_{1},m_{2}} = {1\over m_{2}^{2} - m_{1}^{2}} 
\partial^{\mu}D_{m_{1},m_{2};\mu}
\ee
and the relation (\ref{G3}) is preserved; moreover the order of singularity
is preserved:
$
\omega(D^{adv}_{m_{1},m_{2}}) = \omega(D^{adv}_{m_{1},m_{2};\rho}) + 1 = 0.
$

The Ward identity (\ref{G5}) is non-trivial only for
$M_{1} \not= M_{2}$
and it has the generic form
\be
p(\partial) = c_{0} + c_{2} \partial^{2}.
\ee
If we make the redefinitions:
\be
P^{adv}_{M_{1},M_{2};\mu} \rightarrow P^{adv}_{M_{1},M_{2};\mu}
+ c_{2} \partial_{\mu} \delta, \quad
P^{adv}_{M_{1},M_{2}} \rightarrow P^{adv}_{M_{1},M_{2}}
+ i {c_{0} \over M_{1} - M_{2}} \delta
\ee
the anomaly is eliminated.

It is interesting to summarize the preceding argument by saying that the
anomalies are produced only by the distributions associated to tree graphs.

(iii) It follows that we can describe the structure of the terms from 
$D_{l}^{\mu}(x_{1},x_{2})$
which can produce anomalies. It is sufficient to consider
$l = 1$
and notice that the other piece doubles the value of the anomaly (because of
obvious symmetry properties). We have:
\bea
D^{\mu}_{1}(x_{1},x_{2}) =  
{\partial \over \partial x_{1\mu}} D_{m_{c}}(x_{1}-x_{2})
T_{c}(x_{1},x_{2}) 
+ {\partial^{2} \over \partial x_{1\mu}\partial x^{\rho}_{1}} 
D_{m_{c}}(x_{1}-x_{2}) T_{c}^{\rho}(x_{1},x_{2}) 
\nonumber \\
+ {\partial \over \partial x_{1\mu}} D_{m^{*}_{c}}(x_{1}-x_{2}) 
T^{\prime}_{c}(x_{1},x_{2}) 
+ {\partial^{2} \over \partial x_{1\mu}\partial x^{\rho}_{1}} 
D_{m^{*}_{c}}(x_{1}-x_{2}) T_{c}^{\prime\rho}(x_{1},x_{2}) 
\nonumber \\
+ \sum_{\alpha=1}^{8} :U^{(\alpha)}_{A}(x_{1}) \gamma^{\mu} 
S_{A}(x_{1}-x_{2}) V^{(\alpha)}_{A}(x_{2})
+ \sum_{\alpha=1}^{8} :T^{(\alpha)}_{A}(x_{1}) S_{A}(x_{1}-x_{2}) 
\gamma^{\mu} W^{(\alpha)}_{A}(x_{2}): + \cdots
\eea
where by 
$\cdots$ 
we mean the contributions which do not produce anomalies because of the
argument of (ii). We have the following explicit expressions:
\bea
T_{c}(x_{1},x_{2}) = T_{c}^{YM}(x_{1},x_{2}) 
+ f_{abc} :u_{a}(x_{1}) A_{b}^{\rho}(x_{1}) j_{c\rho}(x_{2}):
\nonumber \\
T^{\prime}_{c}(x_{1},x_{2}) = T_{c}^{\prime YM}(x_{1},x_{2}) 
- f'_{cab} :\Phi_{a}(x_{1}) u_{b}(x_{1}) j_{c}(x_{2}):
\nonumber \\
T_{c}^{\rho}(x_{1},x_{2}) = T_{c}^{YM,\rho}(x_{1},x_{2}),\quad
T_{c}^{\prime\rho}(x_{1},x_{2}) = T_{c}^{\prime YM,\rho}(x_{1},x_{2})
\eea
where the expressions
$T_{c}^{YM}(x_{1},x_{2}), \quad T_{c}^{\prime YM}(x_{1},x_{2})$
and
$T_{c}^{YM,\rho}(x_{1},x_{2}), \quad T_{c}^{\prime YM,\rho}(x_{1},x_{2})$
corresponds to the pure Yang-Mills and can be found in \cite{Gr2}. Also
\bea
U^{(1)}_{A}(x) = U^{(3)}_{A}(x) = U^{(5)}_{A}(x) =  U^{(7)}_{A}(x) \equiv 
(t_{a})_{BA} u_{a}(x) \overline{\psi}_{B}(x), 
\nonumber \\
U^{(2)}_{A}(x) = U^{(4)}_{A}(x) = U^{(6)}_{A}(x) = U^{(8)}_{A}(x) \equiv 
- (t'_{a})_{BA} u_{a}(x) \overline{\psi}_{B}(x) \gamma_{5}, 
\nonumber \\
V^{(1)}_{A}(x)= V^{(4)}_{A}(x) \equiv 
(t_{b})_{AD} \gamma_{\rho} \psi_{D}(x) A_{b}^{\rho}(x),\quad
V^{(2)}_{A}(x) = V^{(3)}_{A}(x) \equiv 
- (t'_{b})_{AD} \gamma_{\rho} \gamma_{5} \psi_{D}(y) A_{b}^{\rho}(x),
\nonumber \\
V^{(5)}_{A}(x)= V^{(8)}_{A}(x) \equiv 
(s_{b})_{AD} \psi_{D}(x) \Phi_{b}(x), \quad
V^{(6)}_{A}(x) = V^{(7)}_{A}(x) \equiv 
(s'_{b})_{AD} \gamma_{5} \psi_{D}(x) \Phi_{b}(x).\qquad 
\label{uv}
\eea
and
\bea
W^{(1)}_{A}(x) = W^{(3)}_{A}(x) = W^{(5)}_{A}(x) =  W^{(7)}_{A}(x) \equiv 
(t_{a})_{AB} u_{a}(x) \psi_{B}(x), 
\nonumber \\
W^{(2)}_{A}(x) = W^{(4)}_{A}(x) = W^{(6)}_{A}(x) = W^{(8)}_{A}(x) \equiv 
(t'_{a})_{BA} u_{a}(x) \gamma_{5} \psi_{B}(x) , 
\nonumber \\
T^{(1)}_{A}(x)= T^{(4)}_{A}(x) \equiv 
- (t_{b})_{CA} \overline{\psi}_{C}(x) \gamma_{\rho} A_{b}^{\rho}(x),\quad
T^{(2)}_{A}(x) = T^{(3)}_{A}(x) \equiv 
- (t'_{b})_{CA} \overline{\psi}_{C}(x) \gamma_{\rho} \gamma_{5} 
A_{b}^{\rho}(x),
\nonumber \\
T^{(5)}_{A}(x) = T^{(8)}_{A}(x) \equiv 
- (s_{b})_{CA} \overline{\psi}_{C}(x) \Phi_{b}(x), \quad
T^{(6)}_{A}(x) = T^{(7)}_{A}(x) \equiv 
- (s'_{b})_{CA}  \overline{\psi}_{C}(x) \gamma_{5} \Phi_{b}(x).\qquad 
\label{wt}
\eea

The expression of the anomaly can be obtained in the generic form:
\be
A(x_{1},x_{2}) = \delta (x_{1} - x_{2}) A(x_{1})
\ee
where:
\bea
A(x_{1}) \equiv
\sum_{c} \left[ T_{c}(x_{1},x_{1}) + T^{\prime}_{c}(x_{1},x_{1}) 
- \left({\partial\over \partial x^{\rho}_{1}}T_{c}^{\rho}\right) (x_{1},x_{1})
- \left({\partial\over \partial x^{\rho}_{1}}
T_{c}^{\prime\rho}\right) (x_{1},x_{1}) 
\right]
\nonumber \\
+ \sum_{\alpha} \left[ :U^{(\alpha)}_{A}(x_{1}) V^{(\alpha)}_{A}(x_{1}):
+ :T^{(\alpha)}_{A}(x_{1}) W^{(\alpha)}_{A}(x_{1}): \right].
\eea

So, the expression of the anomaly 
$A(x)$
gets an extra term because of the presence of the Dirac Fermions:
\bea
A(x) = A^{YM}(x) + i:u_{a}(x) A_{b}^{\rho}(x) \overline{\psi}_{A}(x)
\gamma_{\rho} ([t_{a}, t_{b}] + [t'_{a}, t'_{b}] - i f_{abc} t_{c})_{AB} 
\psi_{B}(x): 
\nonumber \\
+ i:u_{a}(x) A_{b}^{\rho}(x) \overline{\psi}_{A}(x) \gamma_{\rho}  \gamma_{5}
([t_{a}, t'_{b}] + [t'_{a}, t_{b}] - i f_{abc} t'_{c})_{AB} 
\psi_{B}(x): 
\nonumber \\
+ i:u_{a}(x) \Phi_{b}(x) \overline{\psi}_{A}(x) 
([t_{a}, s_{b}] - \{t'_{a}, s'_{b}\} + i f'_{cba} s_{c})_{AB} 
\psi_{B}(x): 
\nonumber \\
+ i:u_{a}(x) \Phi_{b}(x) \overline{\psi}_{A}(x) \gamma_{5}
([t_{a}, s'_{b}] - \{t'_{a}, s_{b}\} + i f'_{cba} s'_{c})_{AB} 
\psi_{B}(x): 
\eea

(iv) We proceed now as in \cite{Gr2}. First we equate the expression
$A(x)$
to a coboundary
$d_{Q} L(x)$.

We get all the relations from \cite{Gr2} (and this explains the first four
relations from the statement). Moreover we obtain for all 
$a,b = 1,\dots,r$:
\bea
[t_{a}, t_{b}] + [t'_{a}, t'_{b}] = i f_{abc} t_{c}, 
\quad
[t_{a}, t'_{b}] + [t'_{a}, t_{b}] = i f_{abc} t'_{c},
\nonumber \\
~[t_{a}, s_{b}] - \{t'_{a}, s'_{b}\} = - i f'_{cba} s_{c},
\quad 
[t_{a}, s'_{b}] - \{t'_{a}, s_{b}\} = - i f'_{cba} s'_{c}.
\eea
which are equivalent to the last two relations from the statement. 

(v) From the preceding computations we can obtain the explicit expression for 
the coboundary
$L(x)$:
it coincides with the expression obtained for the pure Yang-Mills case:
\bea
L(x) = L^{YM}(x) \equiv {1\over 4} f_{cab} f_{cde} 
:A_{a\nu}(x) A_{b\nu}(x) A_{d}^{\mu}(x) A_{e}^{\nu}(x): 
\nonumber \\
- f'_{cda} f'_{ceb} 
:A_{a\nu}(x) A_{b}^{\nu}(x) \Phi_{d}(x) \Phi_{e}(x): 
\nonumber \\
- \sum_{m_{b} \not= 0} g'_{abcd}
:\Phi_{a}(x) \Phi_{b}(x) \Phi_{d}(x) \Phi_{e}(x): 
\label{L}
\eea
where:
\be
g'_{abcd} \equiv {1 \over 2m_{b}} {\cal S}_{abde} f'_{cab} f^{"}_{cde}.
\label{g'}
\ee

Let us also define:
\be
L^{\mu}(x) \equiv \sum_{c} \left[ T^{\mu}_{c}(x,x) 
+ T^{\prime\mu}_{c}(x,x) \right]
\ee

Again it coincides with the expression from the pure Yang-Mills case:
\be
L^{\mu}(x) = L^{YM,\mu}(x) =
- f_{cab} f_{cde} 
:u_{a}(x) A_{b\nu}(x) A_{d}^{\nu}(x) A_{e}^{\mu}(x): 
- f'_{cab} f'_{cde} 
:\Phi_{a}(x) u_{b}(x) \Phi_{d}(x) A_{e}^{\mu}(x):
\ee

We consider now a {\it canonical} causal splitting
$A^{c}(x_{1},x_{2})$
and
$A^{c,\mu}_{l}(x_{1},x_{2})$
given by the expressions which are obtained from the corresponding commutators
if we make the substitutions:
$\Delta \rightarrow \Delta^{adv}$.
This indeed gives a causal splitting of
$D(x_{1},x_{2})$
and
$D^{\mu}_{l}(x_{1},x_{2})$
respectively. However the identity (\ref{gauge-A}) is not fulfilled. 
If we define now the new causal splitting
\be
A(x_{1},x_{2}) \equiv A^{c}(x_{1},x_{2}) + \delta (x_{1} - x_{2}) L(x_{1}),
\quad
A^{\mu}_{l}(x_{1},x_{2}) \equiv A^{c,\mu}_{l}(x_{1},x_{2}) 
+ \delta (x_{1} - x_{2}) L^{\mu}(x_{1})
\ee
then one can see that (\ref{gauge-A}) becomes true. Moreover, in this way one
can obtain in the usual way the expression of the chronological products
$T(x_{1},x_{2})$
and
$T^{\mu}_{l}(x_{1},x_{2})$
such that we have (\ref{gauge}) and all other properties, in particular
symmetry. 
$\qed$
\begin{rem}
If we do not require that (\ref{deg-chrono}) is fulfilled, the relations
(\ref{f'g}) and (\ref{WE}) acquire a weaker form.
\end{rem}

The group-theoretical informations contained in this theorem are:

(a) The expressions
$f_{abc}$
are the structure constants of a Lie algebra $\mathfrak{g}$.

(b) The structure constants
$f_{abc}$
corresponding to
$m_{a} = m_{b} = m_{c} = 0$
generate a Lie subalgebra
$\mathfrak{g}_{0} \subset \mathfrak{g}$.

(c) The 
$r \times r$
(antisymmetric) matrices
$T_{a}, \quad a = 1,\dots,r$
defined according to
\be
(T_{a})_{bc} \equiv - f'_{bca}, \quad \forall a,b,c = 1,\dots,r.
\ee
are an $r$-dimensional representation of the Lie algebra
$\mathfrak{g}$.

The representation
$T_{a}$
exhibited in the statement of the theorem is nothing else but the
representation of the gauge algebra
$\mathfrak{g}$ 
into which the Higgs fields live.

(d) The relation (\ref{repr}) tells that the matrices 
$t_{a}^{\epsilon}$
are representations of the Lie algebra 
$\mathfrak{g}$ 
and relation (\ref{WE}) shows that the matrices
$s_{a}^{\epsilon}$
are some tensor operators with respect to the couple of representations
$t^{\epsilon}_{b}$
of the Lie algebra
$\mathfrak{g}$.          

So, we propose the following strategy of analyzing the generalization of the
standard model described in this paper: first one should find out restrictions
on the Lie algebra 
$\mathfrak{g}$ 
from the relation (\ref{repr-f'}), then one takes a couple of representations 
$t_{a}^{\epsilon}$ 
of this Lie algebra and afterwards one determines the matrices
$s_{a}^{+}$ 
from the relation (\ref{WE}) using ideas from the proof of Wigner-Eckart
theorem.  We mention that if one tries to substitute the formula (\ref{s}) into
the formula (\ref{WE}), as it is done in \cite{AS}, then we end up with some
very complicated trilinear relations which are extremely difficult to analyze
in the general case.

Next, we have a generalization of Proposition 3.9 from \cite{Gr2}. By
definition the {\it Feynman propagator} and the {\it Feynman antipropagator}
are:
\be
\Delta^{F} \equiv \Delta^{adv} - \Delta^{(-)} = \Delta^{ret} + \Delta^{(+)},
\quad
\Delta^{AF} \equiv \Delta^{(+)} - \Delta^{adv} = - \Delta^{ret} - \Delta^{(-)}.
\label{propagator}
\ee

Then we have:
\begin{prop}
Suppose that that there is no contribution 
$T_{1,matter}$
in the first order chronological product. Then, we have
\bea
T^{c}(x,y) = T^{YM,c}(x,y)
\nonumber \\
- f_{abc} D_{m_{c}}^{F}(x-y)
[:A_{a\nu}(x) F_{b}^{\nu\rho}(x) j_{c\rho}(y): - :u_{a}(x)
\partial_{\rho}\tilde{u}_{b}(x) j_{c}^{\rho}(y): + (x \leftrightarrow y)]
\nonumber \\
- f_{abc} {\partial \over \partial x^{\mu}} D_{m_{c}}^{F}(x-y) 
[:A_{a}^{\rho}(x) A_{b}^{\mu}(x) j_{c\rho}(y): - (x \leftrightarrow y)]
\nonumber \\
- f'_{abc} D_{m_{c}}^{F}(x-y)
[:\Phi_{a}(x) \partial_{\mu}\Phi_{b}(x) j_{c}^{\mu}(y): 
- (x \leftrightarrow y)]
\nonumber \\
- f'_{abc} D_{m^{*}_{c}}^{F}(x-y)
[:\partial_{\mu}\Phi_{a}(x) A_{b}^{\mu}(x) j_{c}(y): + (x \leftrightarrow y)] 
\nonumber \\
- f'_{abc} {\partial \over \partial x^{\mu}} D_{m^{*}_{c}}^{F}(x-y)
[:\Phi_{a}(x) A_{b}^{\mu}(x) j_{c}(y): - (x \leftrightarrow y)]
\nonumber \\
- 2h^{(1)}_{abc} D_{m_{c}}^{F}(x-y)
[:\Phi_{a}(x) A_{b}^{\mu}(x) j_{c\mu}(y): + (x \leftrightarrow y)]
\nonumber \\
+ h^{(1)}_{cab} D_{m^{*}_{c}}^{F}(x-y)
[:A_{a\mu}(x) A_{b}^{\mu}(x) j_{c}(y): + (x \leftrightarrow y)]
\nonumber \\
+ h^{(2)}_{cab} D_{m^{*}_{c}}^{F}(x-y)
[:\tilde{u}_{a}(x) u_{b}(x) j_{c}(y): + (x \leftrightarrow y)]
\nonumber \\
+ 3h^{(3)}_{abc} D_{m^{*}_{c}}^{F}(x-y)
[:\Phi_{a}(x) \Phi_{b}(x) j_{c}(y): + (x \leftrightarrow y)]
\nonumber \\
+ 4g_{abcd} D_{m^{*}_{c}}^{F}(x-y)
[:\Phi_{a}(x) \Phi_{b}(x) \Phi_{c}(x) j_{c}(y): + (x \leftrightarrow y)]
\nonumber \\            
+ :A_{a}^{\mu}(x) A_{b}^{\rho}(y): \{[
(t_{a})_{AC} (t_{b})_{CB}
:\overline{\psi}_{A}(x)  \gamma_{\mu} S_{M_{C}}^{F}(x-y) 
\gamma_{\rho} \psi_{B}(y):
\nonumber \\
+ (t'_{a})_{AC} (t'_{b})_{CB}
:\overline{\psi}_{A}(x)  \gamma_{\mu} \gamma_{5} S_{M_{C}}^{F}(x-y) 
\gamma_{\rho} \gamma_{5} \psi_{B}(y):
\nonumber \\
+ (t_{a})_{AC} (t'_{b})_{CB}
:\overline{\psi}_{A}(x)  \gamma_{\mu} S_{M_{C}}^{F}(x-y) 
\gamma_{\rho} \gamma_{5} \psi_{B}(y):
\nonumber \\
+ (t'_{a})_{AC} (t_{b})_{CB}
:\overline{\psi}_{A}(x)  \gamma_{\mu} \gamma_{5} S_{m_{C}}^{F}(x-y) 
\gamma_{\rho} \psi_{B}(y): 
- (a \leftrightarrow b, \mu \leftrightarrow \rho, x \leftrightarrow y) ]
\nonumber \\
+ (t_{a})_{AB} (t_{b})_{BA} P^{F}_{M_{A}M_{B};\mu\rho}(x-y) +
(t'_{a})_{AB} (t'_{b})_{BA} Q^{F}_{M_{A}M_{B};\mu\rho}(x-y) \}
\nonumber \\
+ :\Phi_{a}(x) \Phi_{b}(y): \{[
(s_{a})_{AC} (s_{b})_{CB}
:\overline{\psi}_{A}(x) S_{M_{C}}^{F}(x-y) \psi_{B}(y):
\nonumber \\
+ (s'_{a})_{AC} (s'_{b})_{CB}
:\overline{\psi}_{A}(x)  \gamma_{5} S_{M_{C}}^{F}(x-y) \gamma_{5} \psi_{B}(y):
\nonumber \\
+ (s_{a})_{AC} (s'_{b})_{CB}
:\overline{\psi}_{A}(x)  S_{M_{C}}^{F}(x-y) \gamma_{5} \psi_{B}(y):
\nonumber \\
+ (s'_{a})_{AC} (s_{b})_{CB}
:\overline{\psi}_{A}(x) \gamma_{5} S_{M_{C}}^{F}(x-y) \psi_{B}(y): 
- (a \leftrightarrow b, x \leftrightarrow y) ]
\nonumber \\
+ (s_{a})_{AB} (s_{b})_{BA} P^{F}_{M_{A},M_{B}}(x-y) +
(s'_{a})_{AB} (s'_{b})_{BA} Q^{F}_{M_{A},M_{B}}(x-y) \}
\nonumber \\
+ :A_{a}^{\mu}(x) \Phi_{b}^{\rho}(y): \{[
(t_{a})_{AC} (s_{b})_{CB}
:\overline{\psi}_{A}(x)  \gamma_{\mu} S_{M_{C}}^{F}(x-y) \psi_{B}(y):
\nonumber \\
- (s_{b})_{AC} (t_{a})_{CB}
:\overline{\psi}_{A}(y) S_{M_{C}}^{F}(y-x) \gamma_{\mu} \psi_{B}(x):
\nonumber \\
+ (t'_{a})_{AC} (s'_{b})_{CB}
:\overline{\psi}_{A}(x) \gamma_{\mu} \gamma_{5} S_{M_{C}}^{F}(x-y) \gamma_{5}
\psi_{B}(y):
\nonumber \\
- (s'_{b})_{AC} (t'_{a})_{CB}
:\overline{\psi}_{A}(y) \gamma_{5} S_{M_{C}}^{F}(y-x) \gamma_{\mu} \gamma_{5} 
\psi_{B}(x): 
\nonumber \\
+ (t_{a})_{AC} (s'_{b})_{CB}
:\overline{\psi}_{A}(x)  \gamma_{\mu} S_{M_{C}}^{F}(x-y) 
\gamma_{5} \psi_{B}(y):
\nonumber \\
- (s'_{b})_{AC} (t_{a})_{CB}
:\overline{\psi}_{A}(y) \gamma_{5} S_{M_{C}}^{F}(y-x) 
\gamma_{\mu} \psi_{B}(x): 
\nonumber \\
+ (t'_{a})_{AC} (s_{b})_{CB}
:\overline{\psi}_{A}(x)  \gamma_{\mu}  S_{M_{C}}^{F}(x-y) 
\gamma_{5} \psi_{B}(y):
\nonumber \\
- (s_{b})_{AC} (t'_{a})_{CB}
:\overline{\psi}_{A}(y) \gamma_{5} S_{M_{C}}^{F}(y-x) 
\gamma_{\mu} \psi_{B}(x): 
\nonumber \\
+ (t_{a})_{AB} (s_{b})_{BA} P^{F}_{M_{A},M_{B};\mu}(x-y) +
(t'_{a})_{AB} (s'_{b})_{BA} Q^{F}_{M_{A},M_{B};\mu}(x-y)] 
- [x \leftrightarrow y] \}
\nonumber \\
- D_{m_{a}}^{F}(x-y) :j_{a\mu}(x) j_{a}^{\mu}(y):
\nonumber \\
- (t_{a})_{AC} (t_{a})_{CB} [:\overline{\psi}_{A}(x) \gamma_{\mu}
\Sigma^{F}_{m_{a},M_{C}}(x-y) \gamma^{\mu} \psi_{B}(y): 
+ (x \leftrightarrow y)]
\nonumber \\
- (t'_{a})_{AC} (t'_{a})_{CB} [:\overline{\psi}_{A}(x) \gamma_{\mu} \gamma_{5}
\Sigma^{F}_{m_{a},M_{C}}(x-y) \gamma^{\mu} \gamma_{5} \psi_{B}(y): 
+ (x \leftrightarrow y)]
\nonumber \\
- (t_{a})_{AC} (t'_{a})_{CB} [:\overline{\psi}_{A}(x) \gamma_{\mu}
\Sigma^{F}_{m_{a},M_{C}}(x-y) \gamma^{\mu} \gamma_{5} \psi_{B}(y): 
+ (x \leftrightarrow y)]
\nonumber \\
- (t'_{a})_{AC} (t_{a})_{CB} [:\overline{\psi}_{A}(x) \gamma_{\mu} \gamma_{5}
\Sigma^{F}_{m_{a},M_{C}}(x-y) \gamma^{\mu} \psi_{B}(y): 
+ (x \leftrightarrow y)]
\nonumber \\
- g^{\mu\nu} [(t_{a})_{AB} (t_{a})_{BA} P^{F}_{m_{a};M_{A},M_{B};\mu\nu}(x-y) +
(t'_{a})_{AB} (t'_{a})_{BA} Q^{F}_{m_{a};M_{A},M_{B};\mu\nu}(x-y)]
\nonumber \\
+ D_{m^{*}_{a}}^{F}(x-y) :j_{a}(x) j_{a}(y):
\nonumber \\
+ (s_{a})_{AC} (s_{a})_{CB} [:\overline{\psi}_{A}(x) 
\Sigma^{F}_{m_{a},M_{C}}(x-y) 
\psi_{B}(y): + (x \leftrightarrow y)]
\nonumber \\
+ (s'_{a})_{AC} (s'_{a})_{CB} [:\overline{\psi}_{A}(x) \gamma_{5}
\Sigma^{F}_{m_{a},M_{C}}(x-y) \gamma_{5} \psi_{B}(y): + (x \leftrightarrow y)]
\nonumber \\
+ (s_{a})_{AC} (s'_{a})_{CB} [:\overline{\psi}_{A}(x) 
\Sigma^{F}_{m_{a},M_{C}}(x-y) 
\gamma_{5} \psi_{B}(y): + (x \leftrightarrow y)]
\nonumber \\
+ (s'_{a})_{AC} (s_{a})_{CB} [:\overline{\psi}_{A}(x) \gamma_{5}
\Sigma^{F}_{m_{a},M_{C}}(x-y) \psi_{B}(y): + (x \leftrightarrow y)] 
\label{T2-YM}
\eea
Here 
$
h^{(1)}_{abc} \equiv {1\over 2}(f'_{bca} m_{b} + f'_{acb} m_{a}), \quad
h^{(2)}_{abc} \equiv f'_{abc} m_{b}.
$
\end{prop}
                       
Let us note that the expressions (\ref{vector-current}) and
(\ref{scalar-current}) for the currents can be also written as follows:
\be
j_{a}^{\mu}(x) = 
:\overline{\psi^{+}_{A}}(x) (t_{a}^{+})_{AB} \gamma^{\mu} \psi^{+}_{B}(x): 
+ :\overline{\psi^{-}_{A}}(x) (t_{a}^{-})_{AB} \gamma^{\mu} \psi^{-}_{B}(x):
\label{vector-current-pm}
\ee
and
\be
j_{a}(x) = 
:\overline{\psi^{-}_{A}}(x) (s_{a}^{+})_{AB} \psi^{+}_{B}(x): 
+ :\overline{\psi^{+}_{A}}(x) (s_{a}^{-})_{AB} \psi^{-}_{B}(x):
\label{scalar-current-pm}
\ee
where we have defined for
\be
\psi^{\epsilon}_{A}(x) 
\equiv {1 + \epsilon \gamma_{5} \over 2} \psi_{A}(x),  \quad \epsilon = \pm
\ee
and the components corresponding to the signs $+$ (resp. $-$) are called 
{\it chiral} components of the currents. 

\subsection{The Conservation of the BRST Current}

The expression
\be
j^{\mu}_{BRST}(x) \equiv \left(\partial\cdot A_{a} + m_{a} \Phi_{a} \right)
\stackrel{\leftrightarrow}\partial^{\mu} u_{a}
\ee
is called the {\it BRST current}. One can verify easily the {\it conservation
of the BRST current}:
\be
\partial_{\mu} j^{\mu}_{BRST} = 0.
\ee

Formally, the BRST charge is given by
\be
Q = \int_{\R^{3}} d^{3}x  j^{0}_{BRST}(x).
\ee

We want to investigate the conservation of this current in higher orders of
perturbation theory. We present here the analysis in the second order. First we
have:
\begin{prop}
The following relation is verified:
\bea
[j^{\mu}_{BRST}(x_{1}), T(x_{2})] = 
D_{m_{a}}(x_{1} - x_{2}) A^{\mu}_{a}(x_{1},x_{2}) 
+ \partial^{\mu}D_{m_{a}}(x_{1} - x_{2}) B_{a}(x_{1},x_{2}) 
\nonumber \\
+ \partial^{\rho}D_{m_{a}}(x_{1} - x_{2}) A^{\mu\rho}_{a}(x_{1},x_{2})
+ \partial^{\mu} \partial_{\rho}D_{m_{a}}(x_{1} - x_{2}) 
B^{\rho}_{a}(x_{1},x_{2}) 
\eea
where 
\bea
B_{a}(x_{1},x_{2}) = 
h^{(2)}_{bac} :\partial_{\nu}A^{\nu}_{a}(x_{1}) \Phi_{b}(y) u_{c}(x_{2}):
+ m_{a} h^{(2)}_{bac} :\Phi_{a}(x_{1}) \Phi_{b}(y) u_{c}(x_{2}):
\nonumber \\
- m_{a} f^{\prime}_{abc} :u_{a}(x_{1}) 
\partial_{\rho}\Phi_{b}(x_{2}) A^{\rho}_{c}(x_{2}):
- m_{a} h^{(1)}_{abc} :u_{a}(x_{1}) A_{b\rho}(x_{2}) A^{\rho}_{c}(x_{2}):
\nonumber \\
- m_{a} h^{(2)}_{abc} :u_{a}(x_{1}) \tilde{u}_{b}(x_{2}) u_{c}(x_{2}):
- 3 m_{a} f^{"}_{bca} :u_{a}(x) \Phi_{b}(y) \Phi_{c}(y):
- m_{a} :u_{a}(x_{1}) j_{b}(x_{2}):
\eea
and
\bea
B^{\rho}_{a}(x_{1},x_{2}) = 
f_{bca} :\partial_{\nu}A^{\nu}_{a}(x_{1}) A^{\rho}_{b}(x_{2}) u_{c}(x_{2}):
- m_{a} f_{bca} :\Phi_{a}(x_{1}) A^{\rho}_{b}(x_{2}) u_{c}(x_{2}):
\nonumber \\
+ f_{abc} :u_{a}(x_{1}) A_{b\nu}(x_{2}) F^{\nu\rho}_{c}(x_{2}):
- f_{abc} :u_{a}(x_{1}) u_{b}(x_{2}) \partial^{\rho}\tilde{u}_{c}(x_{2}):
\nonumber \\
+ f^{\prime}_{bca} :u_{a}(x_{1}) \Phi_{b}(x_{2}) 
\partial^{\rho}\Phi_{c}(x_{2}):
+ m_{c} f^{\prime}_{cba} :u_{a}(x_{1}) \Phi_{b}(x_{2}) A^{\rho}_{c}(x_{2}):
+ :u_{a}(x_{1}) j^{\rho}_{a}(x_{2}):
\eea

The expressions for
$
h^{(1)}_{cab}
$
and
$
h^{(2)}_{abc}
$
have been given in the preceding Proposition.
\end{prop}

The computations are long but straightforward. Applying the procedures of the
preceding Subsection we obtain from here:
\begin{prop}
The following expression
\bea
T^{c}(j^{\mu}_{BRST}(x_{1}), T(x_{2})) = 
D^{F}_{m_{a}}(x_{1} - x_{2}) A^{\mu}_{a}(x_{1},x_{2}) 
+ \partial^{\mu}D^{F}_{m_{a}}(x_{1} - x_{2}) B_{a}(x_{1},x_{2})
\nonumber \\
+ \partial^{\rho}D^{F}_{m_{a}}(x_{1} - x_{2}) A^{\mu\rho}_{a}(x_{1},x_{2})
+ \partial^{\mu}\partial_{\rho}D^{F}_{m_{a}}(x_{1} - x_{2}) 
B^{\rho}_{a}(x_{1},x_{2})
\eea
is valid for the canonical chronological product.
\end{prop}

We have the following result which can be interpreted as a conservation of the
BRST current in the second order of perturbation theory:
\begin{thm}
There exists a finite renormalization such that one has the following
conservation law:
\be
{\partial \over \partial x^{\mu}_{1}} T(j^{\mu}_{BRST}(x_{1}), T(x_{2})) = 
i \delta(x_{1} - x_{2}) {\partial \over \partial x^{\mu}_{1}} T^{\mu}(x_{1})
\label{conservation-BRST}
\ee
\end{thm}

{\bf Proof:} We start from the obvious relation
\be
{\partial \over \partial x^{\mu}_{1}} [(j^{\mu}_{BRST}(x_{1}), T(x_{2})] = 0
\ee
and perform the canonical causal splitting using the expression of the
commutator derived above. If we proceed in analogy to the derivation of the
consistency conditions for the second order chronological products we obtain:
\be
{\partial \over \partial x^{\mu}_{1}} T^{c}(j^{\mu}_{BRST}(x_{1}), T(x_{2})) = 
- i {\partial \over \partial x^{\mu}_{1}} 
\left[ \delta(x_{1} - x_{2}) N^{\mu}(x_{1}) \right]
- i \delta(x_{1} - x_{2}) A(x_{1})
\ee
where:
\be
A(x_{1}) \equiv \sum_{a} \left[ B_{a}(x_{1},x_{1})
- \left({\partial B^{\mu}_{a} \over \partial x^{\mu}_{1}}\right)(x_{1},x_{1}) 
\right]
\ee
and
\be
N^{\mu}(x_{1}) \equiv \sum_{a} B^{\mu}_{a}(x_{1},x_{1}).
\ee
Now it is a matter of computation to prove that we have:
$
A = - \partial_{\mu}T^{\mu}.
$
If we perform the finite renormalization
$
T(j^{\mu}_{BRST}(x_{1}), T(x_{2})) = T^{c}(j^{\mu}_{BRST}(x_{1}), T(x_{2}))
-i \delta(x_{1} - x_{2}) N^{\mu}(x_{1})
$
then we obtain the conclusion from the statement.
$\qed$

We remark that one can obtain from the preceding result the gauge invariance
condition (\ref{gauge-inf}) for 
$n = 1$ 
using the method of Appendix B of \cite{DF}.  
\newpage
 
\section{Third-Order Gauge Invariance\label{third}}
                        
\subsection{The Derivation of the Anomaly}
                        
In this Section we will analyze the possible obstructions to factorization of
the $S$-matrix in the third order of the perturbation theory.  In principle,
there is no difference with respect to the preceding Section. Nevertheless, the
details of distribution splitting are considerably more complicated and the
same is true for the whole combinatorial argument. 

First we first give a standard regularization procedure of the distributions
appearing in the lists (\ref{distr0})-(\ref{distr3}). We we choose
$m_{0} > 0$ 
different from all masses of the model and write the Pauli-Villars distribution
for any mass as follows:
\be
D_{m} = D_{m_{0}} + D^{reg};
\label{reg}
\ee
one can check that the order of singularity of
$D^{reg}$
is
\be
\omega(D^{reg}) = - 4.
\ee

The decomposition (\ref{reg}) induces a similar decomposition for all
distributions in the lists (\ref{distr0})-(\ref{distr3}): we have a sum of two
pieces:
\be
\Delta = \Delta^{0} + \Delta^{reg}
\ee
where
\be
\omega(\Delta^{0}) = \omega(\Delta), \quad
\omega(\Delta^{reg}) = \omega(\Delta) - 2
\ee
and the support properties of
$\Delta^{0}$
in the momentum space are more convenient. We have:
\be
\tilde{\Delta}^{0,(\pm)}(p) \sim \theta(\pm p_{0}) f(p^{2})
\ee
with
$supp(f) \subset \{ p^{2} \geq \lambda^{2}\}$
for some parameter with mass significance
$\lambda > 0$;
(for the distributions
$\tilde{\Delta}^{(\pm)}(p)$
we can have in principle
$\lambda = 0$.)

The main result is contained in the following theorem:
\begin{thm}
Suppose that the distribution
$T(x_{1},x_{2},x_{3})$
verifies the condition (\ref{deg-chrono}). Then it verify the formal adiabatic
limit condition if and only if, beside the conditions from the statement of
theorem \ref{T1}, we also have the following set of supplementary conditions:
\be
Tr\left(t^{+}_{a} \{t^{+}_{b},t^{+}_{c}\}\right) =
Tr\left(t^{-}_{a} \{t^{-}_{b},t^{-}_{c}\}\right),
\label{abbj}
\ee
\be
f'_{abc} g'_{bfgh} + f'_{fbc} g'_{bagh} 
+ f'_{gbc} g'_{bafh} + f'_{hbc} g'_{bafg} = 0.
\label{?}
\ee
\label{ano}
\end{thm}
                        
{\bf Proof:}

(i) As before, we will investigate the third order commutators
\be
D(x_{1},x_{2};x_{3})
= [ T(x_{3}), \overline{T}(x_{1},x_{2}) ]
- [ T(x_{1},x_{3}), \overline{T}(x_{2}) ]
- [ T(x_{2},x_{3}), \overline{T}(x_{1}) ]
\label{D3}
\ee
and
\bea
D^{\mu}_{1}(x_{1},x_{2};x_{3})
= [T(x_{3}), \overline{T}^{\mu}_{1}(x_{1},x_{2}) ]
- [T^{\mu}_{1}(x_{1},x_{3}), \overline{T}(x_{2}) ]
- [T(x_{2},x_{3}), \overline{T}^{\mu}_{1}(x_{1}) ],
\nonumber \\
D^{\mu}_{2}(x_{1},x_{2};x_{3})
= [T(x_{3}), \overline{T}^{\mu}_{2}(x_{1},x_{2}) ]
- [T(x_{1},x_{3}), \overline{T}^{\mu}_{1}(x_{2}) ]
- [T^{\mu}_{1}(x_{2},x_{3}), \overline{T}(x_{1}) ],
\nonumber \\
D^{\mu}_{3}(x_{1},x_{2};x_{3})
= [T^{\mu}_{1}(x_{3}), \overline{T}(x_{1},x_{2}) ]
- [T^{\mu}_{2}(x_{1},x_{3}), \overline{T}(x_{2}) ]
- [T^{\mu}_{2}(x_{2},x_{3}), \overline{T}(x_{1}) ].
\label{D3mu}
\eea
                        
All these operator-valued distributions have the causal support property.

(ii) We convene to denote by generically by
\bea
\Delta^{(+)}_{3}(x_{1} - x_{2}) 
= \prod_{i} < \Omega, \phi_{i}(x_{1}) \psi_{i}(x_{2}) \Omega >,
\nonumber \\
\Delta^{(+)}_{1}(x_{2} - x_{3}) 
= \prod_{j} < \Omega, \phi_{j}(x_{2}) \chi_{j}(x_{3}) \Omega >,
\nonumber \\
\Delta^{(+)}_{2}(x_{3} - x_{1}) 
= \prod_{k} < \Omega, \psi_{k}(x_{3}) \chi_{k}(x_{1}) \Omega >
\eea
the distributions appearing in the analysis of the second order perturbation
theory i.e the lists (\ref{distr0})-(\ref{distr3}). They appear with these
three combinations of arguments from various Wick contractions in the preceding
formul\ae~ for the commutators. Here the fields
$
\phi(x_{1}), \quad \psi(x_{2}), \quad \chi(x_{3}))
$
are factors in the Wick monomials of
$
T(x_{1}), \quad T(x_{2}), \quad T(x_{3}) 
$ 
respectively. If Fermi fields are present one has to take into account the
signs induced by the permutation of the non-commuting factors in defining the
associated distributions
$\Delta^{(-)}$.

We have to investigate the types of numerical distributions with causal support
which can appear from the computation of the four commutators. These
distributions will depend only of two variables
$\xi_{1} \equiv x_{1} - x_{3},\quad \xi_{2} \equiv x_{2} - x_{3}$
due to translation invariance.  It convenient to use again a graph theory
terminology. We define a {\it super-line} to be the assemble of lines of a
Feynman graph connecting two vertices. Then the notions of {\it super-tree} and
{\it super-loop} are obvious and we have only such types of graphs.  We give
the generic form of the distributions associated to them.

(a) First we obtain some distributions containing a factor $\delta$ from
commutators containing a factor
$\delta(x-y) L(x)$
or
$\delta(x-y) L^{\mu}(x)$.
In this case we obtain distributions of the type
\be
d(\Delta)(x_{1},x_{2};x_{3}) = \delta(x_{1} - x_{2}) \Delta(x_{2} - x_{3})
\ee
and other permutations of the variables. Here the distribution
$\Delta$
is one from the lists (\ref{distr0})-(\ref{distr3}).

(b) Next, from super-tree graphs we obtain three types of distribution.

(b1) There exists a super-line between $x_{1}$ and $x_{3}$ and a super-line
between $x_{2}$ and $x_{3}$. In this case one obtains distributions of the
form:

\bea
d_{(3)}(x_{1},x_{2};x_{3}) = 
\Delta^{(+)}_{1}(x_{2} - x_{3}) \Delta^{(-)}_{2}(x_{3} - x_{1})
- \Delta^{(-)}_{1}(x_{2} - x_{3}) \Delta^{(+)}_{2}(x_{3} - x_{1})
\nonumber \\
+ \Delta^{F}_{2}(x_{3} - x_{1}) \Delta_{1}(x_{2} - x_{3}) 
- \Delta^{F}_{1}(x_{2} - x_{3}) \Delta_{2}(x_{3} - x_{1})
\label{d3}
\eea

The causal support of this type of distribution can be checked if one derives a
alternative formul\ae.~If:
\be
d^{adv(ret)}_{(3)}(x_{1},x_{2};x_{3}) = 
\Delta^{ret(adv)}_{2}(x_{3} - x_{1}) \Delta^{adv(ret)}_{1}(x_{2} - x_{3})
\label{d3-adv}
\ee
then we have from (\ref{propagator}):
\be
d_{(3)}(x_{1},x_{2};x_{3}) 
= d^{adv}_{(3)}(x_{1},x_{2};x_{3}) - d^{ret}_{(3)}(x_{1},x_{2};x_{3}).
\label{d1-F}
\ee

Moreover, if one uses the expression of the third-order chronological product
(\ref{chronos-n}) one can prove that the distribution of this type is producing
the Feynman propagator
\be
d^{F}_{(3)}(x_{1},x_{2};x_{3}) = 
\Delta^{F}_{2}(x_{3} - x_{1}) \Delta^{F}_{1}(x_{2} - x_{3})
\ee

(b2) There exists a super-line between $x_{1}$ and $x_{2}$ and a super-line
between $x_{1}$ and $x_{3}$. In this case one obtains distributions of the
form:

\bea
d_{(1)}(x_{1},x_{2};x_{3}) = 
\Delta^{(+)}_{2}(x_{3} - x_{1}) \Delta^{(-)}_{3}(x_{1} - x_{2})
- \Delta^{(-)}_{2}(x_{3} - x_{1}) \Delta^{(+)}_{3}(x_{1} - x_{2})
\nonumber \\
+ \Delta^{AF}_{3}(x_{1} - x_{2}) \Delta_{2}(x_{3} - x_{1}) 
- \Delta^{F}_{2}(x_{3} - x_{1}) \Delta_{3}(x_{1} - x_{2})
\label{d1}
\eea

The causal support of this type of distribution can be also checked if one
derives the alternative formul\ae.~ We define:
\be
d^{adv(ret)}_{(1)}(x_{1},x_{2};x_{3}) = 
\Delta^{ret(adv)}_{3}(x_{1} - x_{2}) \Delta^{ret(adv)}_{1}(x_{3} - x_{1})
\ee
and we have as before:
\be
d_{(1)}(x_{1},x_{2};x_{3})
= d^{adv}_{(1)}(x_{1},x_{2};x_{3}) - d^{ret}_{(1)}(x_{1},x_{2};x_{3}).
\ee

If one uses the expression of the third-order chronological product
(\ref{chronos-n}) one can prove that the distribution of this type is producing
the Feynman propagator
\be
d^{F}_{(1)}(x_{1},x_{2};x_{3}) = 
\Delta^{F}_{3}(x_{1} - x_{2}) \Delta^{F}_{2}(x_{3} - x_{1})
\ee

(b3) There exists a super-line between $x_{1}$ and $x_{2}$ and a super-line
between $x_{2}$ and $x_{3}$. In this case one obtains distributions 
$d_{(2)}(x_{1},x_{2};x_{3})$
of the same form as in case (b2) if one makes
$x_{1} \leftrightarrow x_{2}$.

We will denote the distributions associated to super-tree graphs by
$d_{(i)}(\Delta,\Delta')(x_{1},x_{2};x_{3})$
indicating explicitly the distributions in one variable
$\Delta,\quad \Delta'$
from the lists (\ref{distr0})-(\ref{distr3}) involved in the construction. One 
can verify that if the orders of singularity of these distributions are 
$\omega$ and $\omega'$ 
respectively, then:
\be
\omega(d_{(i)}(\Delta,\Delta')) = 4 + \omega + \omega'.
\label{omega1}
\ee

(c) We consider now graphs with a purely Bosonic super-loop. One obtains the
following type of distribution;
\bea
d_{(123)}(x_{1},x_{2};x_{3}) = \Delta^{AF}_{3}(x_{1} - x_{2}) 
\left[ \Delta^{(+)}_{1}(x_{2} - x_{3}) \Delta^{(-)}_{2}(x_{3} - x_{1})
- \Delta^{(-)}_{1}(x_{2} - x_{3}) \Delta^{(+)}_{2}(x_{3} - x_{1}) \right]
\nonumber \\
+ \Delta^{F}_{2}(x_{3}) - x_{1})
\left[ \Delta^{(+)}_{3}(x_{1} - x_{2}) \Delta^{(-)}_{1}(x_{2} - x_{3})
- \Delta^{(-)}_{3}(x_{1} - x_{2}) \Delta^{(+)}_{1}(x_{2} - x_{3}) \right]
\nonumber \\
+ \Delta^{F}_{1}(x_{2} - x_{3}) 
\left[ \Delta^{(+)}_{1}(x_{3} - x_{1}) \Delta^{(-)}_{3}(x_{1} - x_{2})
- \Delta^{(-)}_{1}(x_{3} - x_{1}) \Delta^{(+)}_{3}(x_{1} - x_{2}) \right];
\qquad
\label{d123}
\eea
for a Fermionic super-loop a overall $-1$ sign appears.

The causal support property can be checked by deriving two alternative
formul\ae~: 
\bea
d_{(123)}(x_{1},x_{2};x_{3}) = - \Delta^{ret}_{3}(x_{1} - x_{2}) 
\left[ \Delta^{(+)}_{1}(x_{2} - x_{3}) \Delta^{(-)}_{2}(x_{3} - x_{1})
- \Delta^{(-)}_{1}(x_{2} - x_{3}) \Delta^{(+)}_{2}(x_{3} - x_{1}) \right]
\nonumber \\
+ \Delta^{adv}_{2}(x_{3}) - x_{1})
\left[ \Delta^{(+)}_{3}(x_{1} - x_{2}) \Delta^{(-)}_{1}(x_{2} - x_{3})
- \Delta^{(-)}_{3}(x_{1} - x_{2}) \Delta^{(+)}_{1}(x_{2} - x_{3}) \right]
\nonumber \\
+ \Delta^{adv}_{1}(x_{2} - x_{3}) 
\left[ \Delta^{(+)}_{1}(x_{3} - x_{1}) \Delta^{(-)}_{3}(x_{1} - x_{2})
- \Delta^{(-)}_{1}(x_{3} - x_{1}) \Delta^{(+)}_{3}(x_{1} - x_{2}) \right]
\qquad
\nonumber \\
= - \Delta^{adv}_{3}(x_{1} - x_{2}) 
\left[ \Delta^{(+)}_{1}(x_{2} - x_{3}) \Delta^{(-)}_{2}(x_{3} - x_{1})
- \Delta^{(-)}_{1}(x_{2} - x_{3}) \Delta^{(+)}_{2}(x_{3} - x_{1}) \right]
\nonumber \\
+ \Delta^{ret}_{2}(x_{3}) - x_{1})
\left[ \Delta^{(+)}_{3}(x_{1} - x_{2}) \Delta^{(-)}_{1}(x_{2} - x_{3})
- \Delta^{(-)}_{3}(x_{1} - x_{2}) \Delta^{(+)}_{1}(x_{2} - x_{3}) \right]
\nonumber \\
+ \Delta^{ret}_{1}(x_{2} - x_{3}) 
\left[ \Delta^{(+)}_{1}(x_{3} - x_{1}) \Delta^{(-)}_{3}(x_{1} - x_{2})
- \Delta^{(-)}_{1}(x_{3} - x_{1}) \Delta^{(+)}_{3}(x_{1} - x_{2}) \right].
\qquad
\label{d123-causal}
\eea

We denote suggestively this type of distributions by
$d_{(123)}(\Delta_{1},\Delta_{2},\Delta_{3})(x_{1},x_{2};x_{3})$
where
$\Delta_{i}, \quad i = 1,2,3$
are distributions from the lists (\ref{distr0})-(\ref{distr3}) and we have
concerning the order of singularity:
\be
\omega(d_{(123)}(\Delta_{1},\Delta_{2},\Delta_{3}))
= 4 + \sum_{i}\omega(\Delta_{i}).
\label{omega2}
\ee

We say now something about the generic momentum space structure of such a
distribution. First one has to obtain from the explicit formul\ae~ for the
distributions
$\Delta$
in one variable that in all cases:
\be
\tilde{\Delta_{i}}^{(\pm)}(p) \sim \theta(\pm p_{0}) f_{i}(p^{2})
\ee
with
$supp(f_{i}) \subset \{ p^{2} \geq \lambda_{i}^{2}\}$
for some parameters with mass significance
$\lambda_{i} \geq 0, \quad i = 1,2,3$.
We consider now the Taylor transform of
$\Delta_{(123)}(\xi_{1},\xi_{2})$
and we use the notation 
$K \equiv k_{1} + k_{2}$;
the generic structure is:
\be
\tilde{\Delta}_{(123)}(k_{1},k_{2})
= \theta(k_{1}^{2} - (\lambda_{2} + \lambda_{3})^{2}) g_{1} 
+ \theta(k_{2}^{2} - (\lambda_{3} + \lambda_{1})^{2}) g_{2}
+ \theta(K^{2} - (\lambda_{1} + \lambda_{2})^{2}) g_{3}.
\label{momentum}
\ee

It follows that if at least two of the masses
$\lambda_{i} \geq 0, \quad i = 1,2,3$
are strictly positive, then
$(0,0) \not\in supp(\tilde{\Delta}_{(123)}(k_{1},k_{2})$.
This observation is useful because for causal distributions with such support
property in momentum space one can use the so-called {\it central} formula for
causal decomposition of distributions \cite{Sc1}. If the conditions of validity
of the central formula are not meet we will have to use a regularization
procedure.

(ii) We investigate the possible Ward identities and obstructions to causal
splitting. First we consider the case (b). We illustrate this case on the
the distribution
\be
d^{\mu} \equiv d_{(3)}(\partial^{\mu}D_{m},\Delta)
\ee
where $\Delta$ is arbitrary. The other cases can be treated similarly. First we
derive the Ward identity:
\be
\partial_{\mu}d^{\mu} =
- \delta(x_{1} - x_{3}) \Delta(x_{2} - x_{3})
+ m^{2} D_{m}^{F}(x_{1} - x_{3}) \Delta(x_{2} - x_{3}).
\ee

Using the formula for the causal splitting (\ref{d3-adv}) one can see that the
preceding identity is preserved by the operation of distribution splitting.

Next, we consider case (c). We have to study separately the case when the 
super-loop contains at most one Dirac line and the case when we have three
Dirac lines. We illustrate the first case one the distribution
\be
d^{\mu} = d_{(123)}(\partial^{\mu}D_{m_{1}},D_{m_{2}},D_{m_{3}}), 
\quad m_{1} > 0;
\ee
the other cases can be treated similarly. The Ward identity is in this case:
\be
\partial_{\mu}d^{\mu} = - \delta(x_{2} - x_{3}) D_{m_{2},m_{3}}(x_{3} - x_{1}) 
+ \cdots
\ee
where by $\dots$ we mean contributions with the order of singularity strictly
smaller than $0$.  One computes immediately that both hand sides have the order
of singularity equal to $1$. If we have
$m_{2} + m_{3} > 0$
then we can apply the central decomposition formula and obtain no anomaly. 
In the opposite case, we use standard regularization procedure (\ref{reg}) of
the distributions appearing in the lists (\ref{distr0})-(\ref{distr3})
presented at the beginning of this Subsection.  The decomposition (\ref{reg})
induces a similar decomposition for the distributions of the type 
$d_{(i)}$:
\be
d_{(i)} = d^{0}_{(i)} + d^{reg}
\ee
where
\be
\omega(d^{0}) = \omega(d), \quad \omega(d^{reg}) = \omega(d) - 2
\ee
and the support properties of
$d^{0}$
in the momentum space are more convenient:
$(0,0) \not\in supp(\tilde{d}_{(i)}^{0})$.

If we apply this decomposition to the distributions
$d^{\mu}$
and
$d$
we get two Ward identities, one for each piece.  The first one can be split
causally without anomalies using the central decomposition formula. For the
second identity we note that both hand sides have order of singularity strictly
lower than $-1$ so this relation can be also split causally without anomalies
as explained at the end of Section \ref{perturbation}. In this way we can
obtain a anomaly-free decomposition of the Ward identity we have started with.
One has to check case by case this argument for all the other types of
distributions of type (c) without Dirac loops.

A very important observation is that the preceding argument is not valid for
distributions associated to super-loops containing three Dirac lines. The
reason is that one is lead to the computation of some traces. To be more
specific the relevant terms from the first commutator (\ref{D3mu}) are:
\bea
D^{\mu}_{1}(x_{1},x_{2};x_{3})
= d^{\mu\nu\rho}_{abc}(x_{1},x_{2};x_{3}) 
:u_{a}(x_{1}) A_{b\nu}(x_{2}) A_{c\rho}(x_{3}):
\nonumber \\
+ d^{\mu\nu}_{abc}(x_{1},x_{2};x_{3}) 
:u_{a}(x_{1}) \Phi_{b}(x_{2}) A_{c\nu}(x_{3}): 
+ ( x_{2} \leftrightarrow x_{3} )
\nonumber \\
+ d^{\mu}_{abc}(x_{1},x_{2};x_{3})
:u_{a}(x_{1}) \Phi_{b}(x_{2}) \Phi_{c}(x_{3}):
\nonumber \\
+ d^{\mu}_{a}(x_{1},x_{2};x_{3}) u_{a}(x_{1}) + \cdots
\label{D3-ano}
\eea
where by $\cdots$ we mean the terms which cannot produce anomalies. Let us note
that all these terms are obtain from Wick contractions of the pieces of the
interaction Lagrangian of the type (\ref{dirac}).

The distributions appearing in this formula are {\bf sums} of distributions of
the type
$d_{(123)}$
because of the traces. But in this case, the trace operation can annihilate the
most singular term and instead of (\ref{omega2}) we might have:
\be
\omega(d) < 4 + \sum_{i}\omega(\Delta_{i}).
\label{omega3}
\ee

It follows that these distributions must be studied separately and some
explicit computation are required.

(iii) All the distributions appearing in the formula (\ref{D3-ano}) have $8$
contributions corresponding to the decomposition of the three currents involved
in (\ref{dirac}) into the vectorial and axial components. If we compute the
contribution corresponding to three vectorial factors, then the others can be
obtained by simple substitutions. Let us consider this pure vector contribution
to
$d^{\mu\nu\rho}_{abc}(x_{1},x_{2};x_{3})$;
the explicit expression is:
\be
d^{\mu\nu\rho}_{abc;VVV}
= (t_{a})_{AB} (t_{b})_{BC} (t_{c})_{CA} 
d^{\mu\nu\rho(V)}_{M_{C},M_{A},M_{B}}
+ (t_{a})_{AB} (t_{c})_{BC} (t_{b})_{CA} 
d^{\nu\mu\rho(V)}_{M_{C},M_{B},M_{A}}
\label{VVV}
\ee
where we have defined for arbitrary masses
$M_{1}, M_{2}, M_{3}$
the following fundamental distribution:
\bea
d^{\mu\nu\rho(V)}_{M_{1},M_{2},M_{3}}(x_{1},x_{2};x_{3}) =
\nonumber \\
Tr \{ S^{AF}_{M_{3}}(x_{1} - x_{2}) \gamma^{\nu}
\left[ S_{M_{1}}^{(-)}(x_{2} -x_{3}) \gamma^{\rho} 
S_{M_{2}}^{(+)}(x_{3} - x_{1})
- S_{M_{1}}^{(+)}(x_{2} -x_{3}) \gamma^{\rho} 
S_{M_{1}}^{(-)}(x_{3} - x_{1}) \right]  \gamma^{\mu}
\nonumber \\
+ S^{F}_{M_{1}}(x_{2} - x_{3}) \gamma^{\rho}
\left[ S_{M_{2}}^{(-)}(x_{3} - x_{1}) \gamma^{\mu} 
S_{M_{3}}^{(+)}(x_{1} - x_{2}) 
- S_{M_{2}}^{(+)}(x_{3} - x_{1}) \gamma^{\rho} 
S_{M_{3}}^{(-)}(x_{1} - x_{2}) \right] \gamma^{\nu}
\nonumber \\
+ S^{F}_{M_{2}}(x_{3} - x_{1}) \gamma^{\mu}
\left[ S_{M_{3}}^{(-)}(x_{1} - x_{2}) \gamma^{\nu}
S_{M_{1}}^{(+)}(x_{2} - x_{3}) 
- S_{M_{3}}^{(+)}(x_{1} - x_{2}) \gamma^{\nu}
S_{M_{1}}^{(-)}(x_{2} - x_{3}) \right] \gamma^{\rho} \}
\label{V}
\eea
which is similar to (\ref{d123}); compare also to formula (5.3.11) from
\cite{Sc1}. It also has causal support: one can obtain quite easily alternative
expressions having the structure (\ref{d123-causal}).

The entire vectorial contribution is now obtained if we add the contributions
following from
$d^{\mu\nu\rho}_{abc;VVV}$
if we perform the following simple transforms:
\be
t_{a} \rightarrow t'_{a}, \quad  t_{b} \rightarrow t'_{b}, \quad 
\gamma^{\mu} \rightarrow \gamma^{\mu} \gamma_{5},\quad
\gamma^{\nu} \rightarrow \gamma^{\nu} \gamma_{5},
\ee
and the other two similar possibilities. Using the formula (\ref{5s5}) we
obtain the following form for the pure vectorial part:
\bea
d^{\mu\nu\rho}_{abc;V}
= (t_{a})_{AB} (t_{b})_{BC} (t_{c})_{CA} 
d^{\mu\nu\rho(V)}_{M_{C},M_{A},M_{B}}
+ (t_{a})_{AB} (t_{c})_{BC} (t_{b})_{CA} 
d^{\nu\mu\rho(V)}_{M_{C},M_{B},M_{A}}
\nonumber \\
+ (t'_{a})_{AB} (t'_{b})_{BC} (t_{c})_{CA}
d^{\mu\nu\rho(V)}_{-M_{C},M_{A},M_{B}}
+ (t'_{a})_{AB} (t_{c})_{BC} (t'_{b})_{CA} 
d^{\nu\mu\rho(V)}_{-M_{C},M_{B},M_{A}}
\nonumber \\
+ (t_{a})_{AB} (t'_{b})_{BC} (t'_{c})_{CA} 
d^{\mu\nu\rho(V)}_{M_{C},-M_{A},M_{B}}
+ (t_{a})_{AB} (t'_{c})_{BC} (t'_{b})_{CA} 
d^{\nu\mu\rho(V)}_{M_{C},-M_{B},M_{A}}
\nonumber \\
+ (t'_{a})_{AB} (t_{b})_{BC} (t'_{c})_{CA} 
d^{\mu\nu\rho(V)}_{M_{C},M_{A},-M_{B}}
+ (t'_{a})_{AB} (t'_{c})_{BC} (t_{b})_{CA} 
d^{\nu\mu\rho(V)}_{M_{C},M_{B},-M_{A}}
\label{dV}
\eea

One notices that the vectorial part of
$d^{\mu\nu\rho}_{abc}$
is express only in terms of the distribution of the type
$d^{\mu\nu\rho(V)}_{M_{1},M_{2},M_{3}}$.

By similar transforms one can obtain the pure axial part. One has in defines in
analogy to (\ref{VVV}) the distribution
$d^{\mu\nu\rho(A)}_{M_{1},M_{2},M_{3}}$
by inserting a factor
$\gamma_{5}$:
\be
d^{\mu\nu\rho(A)}_{M_{1},M_{2},M_{3}}(x_{1},x_{2};x_{3}) =
\nonumber \\
Tr \gamma_{5} \left\{ \cdots \right\}
\label{A}
\ee
where by
$\{\cdots\}$
we mean the same paranthesis as in (\ref{V}).  The pure axial contribution to
$d^{\mu\nu\rho}_{abc}$
is similar to (\ref{dV}). The only relevant thing
is that it is expressed only in terms of the new distribution
$d^{\mu\nu\rho(A)}_{M_{1},M_{2},M_{3}}$.
So, it follows that the distribution
$d^{\mu\nu\rho}_{abc}$
can be expressed in terms of two independent distributions:
$d^{\mu\nu\rho(V)}_{M_{1},M_{2},M_{3}}$
and
$d^{\mu\nu\rho(A)}_{M_{1},M_{2},M_{3}}$.
One can prove quite easily that the order of singularities are
\be
\omega(d^{\mu\nu\rho(V)(A)}_{M_{1},M_{2},M_{3}}) = 1.
\label{omega-uAA}
\ee

Let us note in passing that the  asymptotic behaviour of the distribution
\be
d^{\mu\nu\rho}_{abc} = d^{\mu\nu\rho}_{abc;V} + d^{\mu\nu\rho}_{abc;A}
\ee
is given by:
\be
d^{\mu\nu\rho}_{abc} \sim V_{abc} d^{\mu\nu\rho}_{(V)}
+ A_{abc} d^{\mu\nu\rho}_{(A)}
\ee
where the axial tensor
$A_{abc}$
is given by the expression (\ref{ABBJ}) from the Introduction, the vector
tensor is given by a similar expression:
\be
V_{abc} \equiv Tr\left( t^{+}_{a} [ t^{+}_{b}, t^{+}_{c} ]
+ t^{-}_{a} [ t^{-}_{b}, t^{-}_{c} ] \right)
\quad
= \quad f_{bcd} Tr\left( t^{+}_{a} t^{+}_{d} +  t^{-}_{a} t^{-}_{d} \right)
\ee
and
\be
d^{\mu\nu\rho}_{(V)(A)} \equiv d^{\mu\nu\rho(V)(A)}_{0,0,0}.
\ee

A similar investigation can be performed for the other distributions appearing
in the formula (\ref{D3-ano}). The distribution
$d^{\mu\nu}_{abc}$
can be expressed in terms of two independent distributions:
$d^{\mu\nu(V)}_{M_{1},M_{2},M_{3}}$
and
$d^{\mu\nu(A)}_{M_{1},M_{2},M_{3}}$
which can be obtained from
$d^{\mu\nu\rho(V)}_{M_{1},M_{2},M_{3}}$
and
$d^{\mu\nu\rho(A)}_{M_{1},M_{2},M_{3}}$
making
$\gamma^{\rho} \rightarrow 1$.
The order of singularity of these distributions is lower than naive power
counting indicates. They can be written as follows:
\be
d^{\mu\nu(V)(A)}_{M_{1},M_{2},M_{3}} = 
\sum_{i=1}^{3} M_{i} d^{\mu\nu(V)(A)}_{i}
\label{decrease}
\ee
with
\be
\omega(d^{\mu\nu(V)(A)}_{i}) = 0.
\label{omega-uAP}
\ee

Analogously, the distribution
$d^{\mu}_{abc}$
can be expressed in terms of two independent distributions:
$d^{\mu(V)}_{M_{1},M_{2},M_{3}}$
and
$d^{\mu(A)}_{M_{1},M_{2},M_{3}}$
which can be obtained from
$d^{\mu\nu(V)}_{M_{1},M_{2},M_{3}}$
and
$d^{\mu\nu(A)}_{M_{1},M_{2},M_{3}}$
making
$\gamma^{\nu} \rightarrow 1$.
We also have:
\be
\omega(d^{\mu(V)(A)}_{M_{1},M_{2},M_{3}}) = 1.
\label{omega-uPP}
\ee

Finally, the distribution
$d^{\mu}_{a}$
can be expressed in terms of two independent distributions:
$d^{\mu(V)}_{m;M_{1},M_{2},M_{3}}$
and
$d^{\mu(A)}_{m;M_{1},M_{2},M_{3}}$
which can be obtained from
$d^{\mu(V)}_{M_{1},M_{2},M_{3}}$
and
$d^{\mu(A)}_{M_{1},M_{2},M_{3}}$
by making
$S_{M_{1}} \rightarrow \Sigma_{m,M_{1}}$.
We have in this case:
\be
\omega(d^{\mu(V)(A)}_{m;M_{1},M_{2},M_{3}}) = 3.
\label{omega-u}
\ee

We also need the distributions
$d^{(V)}_{M_{1},M_{2},M_{3}}$
and
$d^{(V)}_{m;M_{1},M_{2},M_{3}}$
which can be obtained from
$d^{\mu(V)}_{M_{1},M_{2},M_{3}}$
and
$d^{\mu(V)}_{m;M_{1},M_{2},M_{3}}$
respectively by making
$\gamma^{\mu} \rightarrow 1$. 
In this case we have a structure similar to (\ref{decrease}) and a similar
result for the order of singularity.

We can easily see that all distributions
$d^{\dots (A)}_{\dots}$
are completely antisymmetric in the Lorentz indices due to traces involving a
$\gamma_{5}$. 
matrix. It is not difficult to prove that one can impose a supplementary
condition on the causal splitting procedure, namely the preservation of this
symmetry property.

The distributions appearing into the third commutator from (\ref{D3mu}) can be
obtained from the preceding ones by making the substitution
$x_{1} \leftrightarrow x_{2}$
and this doubles the value of the possible anomalies coming from the first
commutator.

The distributions appearing in to second commutator from (\ref{D3mu}) can be
obtained from the preceding ones by more subtle transforms. For the case (b)
and the case (c) without Dirac loops we have the same list of distributions and
there are no anomalies. For the case (c) with Dirac loops we have to consider:
\bea
d^{\mu\nu\rho}_{M_{1},M_{2},M_{3}}\rightarrow 
d^{\rho\nu\mu}_{M_{1},M_{2},M_{3}}
\nonumber \\
d^{\mu\nu}_{M_{1},M_{2},M_{3}}(x_{1},x_{2};x_{3}) \rightarrow 
f^{\mu\nu}_{M_{2},M_{3},M_{1}}(x_{1},x_{2};x_{3}) \equiv
d^{\nu\mu}_{M_{2},M_{3},M_{1}}(x_{2},x_{3};x_{1})
\nonumber \\
d^{\mu}_{M_{1},M_{2},M_{3}}(x_{1},x_{2};x_{3}) \rightarrow 
f^{\mu}_{M_{1},M_{2},M_{3}}(x_{1},x_{2};x_{3}) \equiv
d^{\nu\mu}_{M_{3},M_{1},M_{2}}(x_{3},x_{1};x_{2}).
\eea

(iv) Now we can give the list of Ward identities verified by these
distributions. Using Dirac equation for the propagators we get:
\bea
i {\partial \over \partial x^{\mu}_{1}} 
d^{\mu\nu\rho(V)}_{M_{1},M_{2},M_{3}}(x_{1},x_{2};x_{3}) =
(M_{1} - M_{3}) f^{\nu\rho(V)}_{M_{1},M_{2},M_{3}}(x_{1},x_{2};x_{3})
\nonumber \\
- i\delta(x_{1}  - x_{2}) P^{\nu\rho}_{M_{2},M_{3}}(x_{1} - x_{3})
- i\delta(x_{1}  - x_{3}) P^{\rho\nu}_{M_{1},M_{2}}(x_{2} - x_{3})
\label{g1}
\eea
\bea
i {\partial \over \partial x^{\mu}_{3}} 
d^{\rho\nu\mu(V)}_{M_{1},M_{2},M_{3}}(x_{1},x_{2};x_{3}) =
(M_{3} - M_{2}) d^{\rho\nu(V)}_{M_{1},M_{2},M_{3}}(x_{1},x_{2};x_{3})
\nonumber \\
+ i\delta(x_{1}  - x_{3}) P^{\rho\nu}_{M_{1},M_{2}}(x_{2} - x_{3})
+ i\delta(x_{2}  - x_{3}) P^{\rho\nu}_{M_{3},M_{1}}(x_{1} - x_{3})
\label{g2}
\eea
\bea
i {\partial \over \partial x^{\mu}_{1}} 
d^{\mu\nu(V)}_{M_{1},M_{2},M_{3}}(x_{1},x_{2};x_{3}) =
(M_{1} - M_{3}) d^{\nu(V)}_{M_{2},M_{3},M_{1}}(x_{2},x_{3};x_{1})
\nonumber \\
- i\delta(x_{1}  - x_{2}) P^{\nu}_{M_{3},M_{2}}(x_{1} - x_{3})
- i\delta(x_{1}  - x_{3}) P^{\nu}_{M_{1},M_{2}}(x_{2} - x_{3})
\label{g3}
\eea
\bea
i {\partial \over \partial x^{\mu}_{3}} 
f^{\mu\nu(V)}_{M_{1},M_{2},M_{3}}(x_{1},x_{2};x_{3}) =
(M_{3} - M_{2}) d^{\nu(V)}_{M_{2},M_{3},M_{1}}(x_{2},x_{3};x_{1})
\nonumber \\
+ i\delta(x_{1}  - x_{3}) P^{\nu}_{M_{1},M_{2}}(x_{2} - x_{3})
+ i\delta(x_{2}  - x_{3}) P^{\nu}_{M_{1},M_{3}}(x_{1} - x_{3})
\label{g4}
\eea
\bea
i {\partial \over \partial x^{\mu}_{1}} 
d^{\mu(V)}_{M_{1},M_{2},M_{3}}(x_{1},x_{2};x_{3}) =
(M_{1} - M_{3}) d^{(V)}_{M_{2},M_{3},M_{1}}(x_{1},x_{2};x_{3})
\nonumber \\
+ i\delta(x_{1}  - x_{3}) P_{M_{1},M_{2}}(x_{2} - x_{3})
+ i\delta(x_{2}  - x_{3}) P_{M_{1},M_{3}}(x_{1} - x_{3})
\label{g5}
\eea
\bea
i {\partial \over \partial x^{\mu}_{1}} 
d^{\mu(V)}_{m;M_{1},M_{2},M_{3}}(x_{1},x_{2};x_{3}) =
(M_{1} - M_{3}) d^{(V)}_{m;M_{2},M_{3},M_{1}}(x_{1},x_{2};x_{3})
\nonumber \\
+ i\delta(x_{1}  - x_{3}) P_{m;M_{1},M_{2}}(x_{2} - x_{3})
+ i\delta(x_{2}  - x_{3}) P_{m;M_{1},M_{3}}(x_{1} - x_{3}).
\label{g5a}
\eea

The Ward identities for the axial distributions present a notable difference.
Because of the trace operation, the delta terms disappear. Using also formula
(\ref{5s5}) we get:
\be
i {\partial \over \partial x^{\mu}_{1}} 
d^{\mu\nu\rho(A)}_{M_{1},M_{2},M_{3}}(x_{1},x_{2};x_{3}) =
(M_{1} + M_{3}) f^{\nu\rho(A)}_{M_{1},M_{2},M_{3}}(x_{1},x_{2};x_{3})
\label{g6}
\ee
\be
i {\partial \over \partial x^{\mu}_{3}} 
d^{\rho\nu\mu(A)}_{M_{1},M_{2},M_{3}}(x_{1},x_{2};x_{3}) =
(M_{3} + M_{2}) d^{\rho\nu(A)}_{M_{1},M_{2},M_{3}}(x_{1},x_{2};x_{3})
\label{g7}
\ee
\be
i {\partial \over \partial x^{\mu}_{1}} 
d^{\mu\nu(A)}_{M_{1},M_{2},M_{3}}(x_{1},x_{2};x_{3}) =
(M_{1} + M_{3}) d^{\nu(A)}_{M_{2},M_{3},M_{1}}(x_{2},x_{3};x_{1})
\label{g8}
\ee
\be
i {\partial \over \partial x^{\mu}_{3}} 
f^{\mu\nu(A)}_{M_{1},M_{2},M_{3}}(x_{1},x_{2};x_{3}) =
(M_{3} + M_{2}) d^{\nu(A)}_{M_{2},M_{3},M_{1}}(x_{2},x_{3};x_{1})
\label{g9}
\ee
\be
i {\partial \over \partial x^{\mu}_{1}} 
d^{\mu(A)}_{M_{1},M_{2},M_{3}}(x_{1},x_{2};x_{3}) = 0
\label{g10}
\ee
\be
i {\partial \over \partial x^{\mu}_{1}} 
d^{\mu(A)}_{m;M_{1},M_{2},M_{3}}(x_{1},x_{2};x_{3}) = 0.
\label{g10a}
\ee

The causal splitting of these two type of Ward identities is sensibly
different. Let us first consider only the first six equations (the vectorial
Ward identities). Because of the delta terms in the right hand sides, we have
the same order of singularity for both sides in all vectorial Ward identities ,
if the conditions of application of the central splitting formula are meet we
get no anomalies. If some of the masses are null, one has to use a
regularization procedure like for the case (b). More precisely one can prove
that the decomposition (\ref{reg}) induces a similar decomposition for the
distributions of the type 
$d_{(123)}$:
\be
d_{(123)} = d^{0}_{(123)} + d^{reg}_{(123)}
\ee
where
\be
\omega(d^{0}_{(123)}) = \omega(d_{(123)}), 
\quad \omega(d^{reg}_{(123)}) = \omega(d_{(123)}) - 4
\ee
and the support properties of
$d^{0}_{(123)}$
in the momentum space are more convenient:
$(0,0) \not\in supp(\tilde{d}_{(123)}^{0})$.

We turn now to the last six equations (the axial Ward identities). The previous
argument is still valid for the last two of them. We consider some generic
anomalies
$P_{1}^{\nu},\quad P^{\nu}_{3}$
obtained after the causal splitting of the identities (\ref{g8}) and
(\ref{g9}). If we differentiate the corresponding equations with respect to
$x^{\nu}_{3}$
and
$x^{\nu}_{1}$
respectively, we obtain the consistency equations
\be
{\partial \over \partial x^{\nu}_{3}} P^{\nu}_{1} = 0, \quad
{\partial \over \partial x^{\nu}_{1}} P^{\nu}_{3} = 0
\ee
and this leads to
$P^{\mu}_{i} = 0, \quad i = 1,3.$

The Ward identities (\ref{g6}) and (\ref{g7}) can produce anomalies of the type
\be
P^{\nu\rho}(X) = {\rm const} \quad\varepsilon^{\nu\rho\alpha\beta} 
{\partial^{2} \over \partial x^{\alpha}_{1} \partial x^{\beta}_{2}} \delta(X)
\ee
for some positive constant 
${\rm const}$. 
The explicit expression of this constant can be computed as in \cite{Sc1}
Section 5.3. One cannot eliminate such type of anomalies from both equations by
redefinitions. The resulting anomaly is then:
\be
A(X) = {\rm const.} \quad A_{abc} \varepsilon^{\nu\rho\alpha\beta} 
{\partial^{2} \over \partial x^{\alpha}_{1} \partial x^{\beta}_{2}} \delta(X)
:u_{a}(x_{1}) A_{b\nu}(x_{2}) A_{c\rho}(x_{3}):
\ee
where:
\be
A_{abc} \equiv 2 \quad Tr\left(t'_{a} \{t'_{b},t'_{c}\} +
t_{a} \{t_{b},t'_{c}\} + t'_{a} \{t_{b},t_{c}\} 
+ t_{a} \{t'_{b},t'_{c}\} \right).
\ee

Performing some redefinitions of the expressions
$A^{\mu}_{l}(X)$
(``integration by parts") we can reexpress this {\it axial anomaly} in the
following form:
\be
A_{ABBJ}(X) = {\rm const.} \quad A_{abc} 
\varepsilon_{\mu\nu\rho\sigma} \delta(X)
:u_{a}(x_{1}) F_{b}^{\mu\nu}(x_{1}) F_{c}^{\rho\sigma}(x_{3}):
\label{axial-ano}
\ee
and we can also show that the tensor depending only on the group indices
$A_{abc}$
is in fact given by the formula (\ref{ABBJ}) from the introduction. 
The anomaly
$A_{ABBJ}$
is a cocycle
\be
d_{Q} A_{ABBJ} = 0
\ee
but it is not a coboundary; so disappears {\it iff} we have the condition
$A_{abc} = 0$
i.e. the well-known condition (\ref{abbj}) from the statement.

(v) We still have to investigate the possible anomalies coming from the delta
terms i.e from distributions associated to graphs of type (a). We present here
briefly the analysis of these terms. One can compute the commutators and select
the terms which will lead, in principle, to an anomaly. We get:
\bea
[T_{1}^{\mu}(x), L(y)] = f_{abc} f_{dcf} f_{dgh} 
{\partial \over \partial x_{\mu}} D_{m_{c}}(x - y) :u_{a}(x) A_{b\nu}(x)
A_{f\lambda}(y) A_{g}^{\nu}(y) A_{h}^{\lambda}(y):
\nonumber \\
- 2 f_{abc} f'_{dec} f'_{dgh} {\partial \over \partial x_{\mu}}
D_{m_{c}}(x - y) :u_{a}(x) A_{b\rho}(x) A_{h}^{\rho}(y) \Phi_{e}(y)
\Phi_{g}(y):
\nonumber \\
+ 2 f'_{abc} f'_{dbf} f'_{dgh} {\partial \over \partial x_{\mu}}
D_{m^{*}_{b}}(x - y) :\Phi_{a}(x) u_{c}(x) A_{f\rho}(y) A_{h}^{\rho}(y)
\Phi_{g}(y):
\nonumber \\
+ 4 f'_{abc} g'_{bfgh}
{\partial \over \partial x_{\mu}} D_{m^{*}_{b}}(x - y) 
:\Phi_{a}(x) u_{c}(x) \Phi_{f}(y) \Phi_{g}(y) \Phi_{h}(y):
+ \cdots
\eea

By $\cdots$ we mean the rest of the commutator which cannot produce anomalies
Now, as in \cite{Gr1} and \cite{Gr2} we get from this commutator a possible
anomaly:
\be
A(x_{1},x_{2},x_{3}) = \delta(X) 
\left[A_{1}(x_{1}) + A_{2}(x_{1}) + A_{3}(x_{1}) + A_{4}(x_{1}) \right]
\ee
where
\be
A_{1}(x) = i f_{abc} f_{dcf} f_{dgh} 
:u_{a}(x) A_{b\nu}(x) A_{f\lambda}(x) A_{g}^{\nu}(x) A_{h}^{\lambda}(x):
\ee 
\be
A_{2}(x) = - 2i f_{abc} f'_{dec} f'_{dgh}
:u_{a}(x) A_{b\rho}(x) A_{h}^{\rho}(x) \Phi_{e}(x) \Phi_{g}(x):
\ee
\be
A_{3}(x) = 2i f'_{abc} f'_{dbf} f'_{dgh} 
:\Phi_{a}(x) u_{c}(x) A_{f\rho}(x) A_{h}^{\rho}(x) \Phi_{g}(x):
\ee
\bea
A_{4}(x) = i\left[ f'_{abc} g'_{bfgh} + f'_{fbc} g'_{bagh} 
+ f'_{gbc} g'_{bafh} + f'_{hbc} g'_{bafg} \right]
\nonumber \\
:\Phi_{a}(x) u_{c}(x) \Phi_{f}(y) \Phi_{g}(x) \Phi_{h}(x):
\eea

The results are:
\begin{itemize}
\item
In \cite{DHKS1} it is proved that the 
$A_{1} = 0$
due to the Jacobi identity. 
\item
One can also show, using the identity (\ref{f-f'})
that 
$A_{2} + A_{3} = 0$.
\item
If we try to write the anomaly
$A_{4}$
as a coboundary
$d_{Q}L(x)$
we should take
\be
L(x) = g'_{acfgh}
:\Phi_{a}(x) \Phi_{c}(x) \Phi_{f}(x) \Phi_{g}(x) \Phi_{h}(x):
\ee
which is forbidden by the assumption that (\ref{deg-chrono}) is fulfilled.  
\end{itemize}
So we get the second restriction from the statement.
$\qed$

\begin{rem}
Recently \cite{HS} a new method was proposed to solve problems of consistency
as the ones appearing in our paper. Instead of imposing a factorisation
condition of the type (\ref{gi-Tn}) (or its ``infinitesimal" version 
(\ref{gauge-inf})) one imposes a quantum analogue of the Noether
conservation law of a certain current. Presumably, this starting points are
equivalent and they should lead to the same sets of consistency conditions.
This point deserves further investigations. However, one should compare
carefully the relation (\ref{conservation-BRST}) expressing the conservation
law of the BRST current (and equivalent to the formal adiabatic limit
condition) to the relation (4.5) of \cite{HS} expression the quantum Noether
postulate.
\end{rem}
                        
%\newpage
\subsection{The Standard Model}

We remind the notations from \cite{Gr2}. The Lie algebra is in this case
$su(2) \times u(1)$
and the standard basis 
$X_{a}, \quad a = 0,1,2,3$
has the usual commutation relations
\be
[ X_{a}, X_{b}] = \epsilon_{abc} X_{c}, \quad a, b = 1,2,3, \quad
[ X_{0}, X_{a}] = 0, \quad a = 1,2,3.
\ee

In the new basis 
$Y_{a}, \quad a = 0,1,2,3$
defined by
\be
Y_{a} = g~X_{a}, \quad a = 1,2, \quad
Y_{3} = - g~cos \theta~ X_{3} + g'~sin~\theta X_{0}, \quad
Y_{0} = - g~sin \theta~ X_{3} - g'~cos~\theta X_{0}.
\ee
(here the angle $\theta$, determined by the condition
$cos~\theta > 0$
is the {\it Weinberg angle} and the constants $g$ and $g'$ are real with 
$g > 0$)
the structure constants are:
\be
f_{210} = g~sin~\theta, \quad f_{321} = g~cos~\theta, \quad 
f_{310} = 0, \quad f_{320} = 0
\label{standard-const}
\ee
and the rest of the constants are determined by antisymmetry. The choice of the
masses is:
\be
m_{0} = 0, \quad m_{a} \not= 0, \quad a = 1,2,3
\ee
(the particles created by
$A^{\mu}_{0}$
being the {\it photons} and the particles created by
$A^{\mu}_{a}, \quad a = 1,2,3$
the {\it heavy Bosons}).

In \cite{Gr2} we have found out the following result: 
\begin{thm}
In the standard model, the following relations are true:

(a) the masses of the heavy Bosons are constrained by:
\be
m_{1} = m_{2} = m_{3} cos~\theta;
\ee

(b) the constants
$f'_{abc}$
are completely determined by the antisymmetry property (\ref{anti-f'}) and:
\be
f'_{011} = f'_{022} = {\epsilon~g\over 2}, \quad
f'_{033} = {\epsilon~g \over 2cos~\theta}, \quad
f'_{210} = g~sin~\theta, \quad
f'_{321} = - f'_{312} = {g \over 2}, \quad
f'_{123} = - g~{cos~2\theta \over 2cos~\theta},
\ee
the rest of them being zero. Here 
$\epsilon$
can take the values $+$ or $-$.

(c) the constants
$f^{"}_{abc}$
are (partially) determined by:
\begin{eqnarray}
f^{"}_{abc} = 0 \quad (a,b,c = 1,2,3), \quad
f^{"}_{001} = f^{"}_{002} = f^{"}_{003} = 
f^{"}_{012} = f^{"}_{023} = f^{"}_{031} = 0, 
\nonumber \\
f^{"}_{011} = f^{"}_{022} = f^{"}_{033} = 
{\epsilon g\over 12 m_{1}} (m_{0}^{H})^{2}. 
\end{eqnarray}

Moreover, one can fix
$\epsilon = +$.
\end{thm}
\begin{rem}
In \cite{Sc2} a dual point of view is followed: one gives the masses of the
heavy Bosons
$m_{1} = m_{2} \not= m_{3}$
and determines that the gauge algebra must be
$su(2) \times u(1)$.
\end{rem}

We consider the minimal standard model containing only one generation of Dirac
particles. In this case one takes in the generic formalism from the preceding
Section:
$N = 2$
and
\be
M \equiv \left( \matrix{ 0 & 0 \cr 0 & m_{e}} \right).
\ee

The components 
$\psi_{2}$
(resp.
$\psi_{1}$)
correspond to the {\it electron} (resp. the electronic {\it neutrino}) and
$m_{e}$
is the electron mass. Remark that the neutrino mass is considered null.

The choice for the representations
$t^{\pm}_{a}$
is the following one:
\bea
t^{+}_{1} = {1\over 2} g \sigma_{1}, \quad 
t^{+}_{3} = {1\over 2}
\left( -g cos~\theta \sigma_{3} + g' sin~\theta {\bf 1}\right), 
\nonumber \\
t^{+}_{2} = {1\over 2} g \sigma_{2}, \quad
t^{+}_{0} = - {1\over 2}
\left(g sin~\theta \sigma_{3} + g' cos~\theta {\bf 1}\right)
\eea
and
\be
t^{+}_{1} = t^{+}_{2} =  0, \quad 
t^{+}_{3} = y~ sin~\theta, \quad t^{+}_{0} = -y~  cos~\theta;
\ee
here 
$\sigma_{i}$
are the Pauli matrices.  The representation property (\ref{repr}) is fulfilled
for any matrix $y$.  However, we have the following elementary result:
\begin{prop}
The interaction between the Dirac field of the electron 
$\psi_{2}$
and the electromagnetic field
$A_{0}^{\mu}$
has the usual form
$$e :\bar{\psi}_{2} \gamma_{\mu} \psi_{2}: A_{0}^{\mu}$$
(here $e$ is the electron charge) {\it iff}
\be
g = {e\over sin~\theta}, \quad g' = - {e\over cos~\theta}
\ee
and
\be
y = {1\over 2} g' ({\bf 1} - \sigma_{3}) = {1\over 2m_{e}} g' M.
\ee
\end{prop}

Next, we have
\begin{prop}
The expressions for the matrices
$s_{a}^{+}$
are:
\be
s_{0}^{+} = {m_{e} \over 2m_{1}} \left( \matrix{ 0 & 0 \cr 0 & 1} \right),
\quad
s_{1}^{+} = {im_{e} \over 2m_{1}} \left( \matrix{ 0 & 0 \cr 1 & 0} \right),
\quad
s_{2}^{+} = - {m_{e} \over 2m_{1}} \left( \matrix{ 0 & 0 \cr 1 & 0} \right),
\quad
s_{3}^{+} = {im_{e} \over 2m_{1}} \left( \matrix{ 0 & 0 \cr 0 & 1} \right).
\ee
\end{prop}

{\bf Proof:}

One uses the relations (\ref{s}) for
$a = 1,2,3$ 
and obtains the expressions for 
$s_{a}^{+}, \quad a = 1,2,3.$
Next, we use the relation (\ref{WE}), more precisely:
\be
t_{a}^{-} s_{0}^{+} - s_{0}^{+} t_{a}^{+} = i f'_{0ca} s_{c}^{+}, \quad
a = 1,2,3.
\ee

This equation gives immediately the expression for
$s^{+}_{0}$.
$\qed$

The expression of the Higgs potential is obtained as in \cite{DS}, \cite{ASD3}.
One can check that in this way the usual standard model is obtained.

%\newpage                        
\subsection{Regularization and Anomalies}

We have succeeded to give a complete analysis of the possible anomalies
appearing in the standard model up to the order three of the
perturbation theory. One would want to generalise this analysis to all
orders of the perturbation theory. It is possible that one can use the same
type of combinatorial argument, namely one considers possible distributions
appearing in the commutators
$D(X)$
of order $n$ and observes that only the super-loop graphs with Dirac lines can
produce anomalies. Then it is quite possible that in higher orders the orders
of singularity are sufficiently lower to make possible a causal splitting of
the Ward identities without anomalies. This seems to be indicated by the
traditional argument from the literature \cite{Ad}, \cite{AB}, \cite{Sch1}.

\end{document}